\documentclass[preprint,12pt,authoryear]{elsarticle}

\usepackage{amssymb}
\usepackage{amsmath}

\usepackage{graphicx}
\usepackage{caption}
\usepackage{subcaption}
\usepackage{rotating}
\usepackage{xcolor}
\usepackage{natbib}
\usepackage{aas_macros}
\usepackage{hyperref}

\journal{Journal of High Energy Astrophysics}

\begin{document}

\begin{frontmatter}



\title{Morphology of Young Massive Stellar Clusters \\with Next-Generation IACTs} 


\author{A. Bonollo\textsuperscript{a,}\textsuperscript{b,}\textsuperscript{c}}
\author{P. Esposito\textsuperscript{a,}\textsuperscript{c}}
\author{A. Giuliani\textsuperscript{c}} 
\author{P. Caraveo\textsuperscript{c}} 
\author{G. Galanti\textsuperscript{c}} 
\author{S. Crestan\textsuperscript{c}} 
\author{M. Rigoselli\textsuperscript{d,}\textsuperscript{c}} 
\author{S. Mereghetti\textsuperscript{c}} 

\affiliation{organization={Scuola Universitaria Superiore IUSS Pavia},
            addressline={Piazza della Vittoria 15}, 
            city={Pavia},
            postcode={27100}, 
            country={Italy}}

\affiliation{organization={Dipartimento di Fisica, Università degli Studi di Trento},
            addressline={Via Sommarive 14}, 
            city={Povo (TN)},
            postcode={38123}, 
            country={Italy}}

\affiliation{organization={INAF -- Istituto di Astrofisica Spaziale e Fisica Cosmica (IASF) di Milano},
            addressline={Via A. Corti 12}, 
            city={Milan},
            postcode={20133}, 
            country={Italy}}

\affiliation{organization={INAF -- Osservatorio Astronomico di Brera},
            addressline={Via Brera 28}, 
            city={Milan},
            postcode={20131}, 
            country={Italy}}

\begin{abstract}
The term PeVatron designates astrophysical objects capable of accelerating particles to PeV energies (1 PeV = $10^{15}$ eV). Their nature and particle acceleration mechanisms are uncertain, but ultra-high-energy gamma rays ($>$100 TeV) are produced when particles accelerated by either leptonic and hadronic PeVatrons interact with the surrounding medium or radiation fields. The atmospheric air shower observatory LHAASO detected photons with energies above 100 TeV from 43 sources in the Galactic Plane, proving the existence of PeVatrons within the Milky Way. In particular, one of the detections was a 1.4 PeV photon in spatial correspondence with Cygnus OB2, providing a strong hint that young massive stellar clusters (YMSCs) can act as PeVatrons.\\
The next-generation ground-based Cherenkov telescopes will have unprecedented energy and angular resolution. Therefore, they will be able to resolve spatially YMSCs better than LHAASO. We focused on a sample of 5 YMSCs and their environments visible from either hemisphere with the CTAO or ASTRI Mini-Array. We modeled the secondary gamma-ray emission above 1 TeV and simulated observations of all sources. We devised methods for classifying YMSCs that could be detected as unidentified extended TeV sources and estimate the observational time needed to distinguish the morphology of different classes of sources.\\
We study the morphology of the sources in our sample in order to identify their main features. We simulated observations of all  sources with the instrument response function (IRF) of CTAO or ASTRI Mini-Array. We compare their emission distribution to the one of the TeV halos observed by HAWC. We parametrize their radial profiles in order to develop methodologies to classify them and to distinguish YMSCs from TeV halos based on their morphology. We expect some feature, such as the emission peak, to be key in differentiating between the two classes of objects. We then test them on a sample of sources of the first LHAASO catalog.
\end{abstract}




\end{frontmatter}



\section{Introduction}
\label{Intro}

The origin of ultra-high-energy (UHE) Galactic cosmic rays (CRs) is still a subject of debate. The general consensus is that the locally detected CR protons and nuclei below the 'knee' at around 1 PeV of the CR spectrum are produced in the Milky Way \citep{CR_review}. Supernova remnant (SNR) shocks are considered the main contributors up to the spectral knee, but they are widely reckoned to have difficulties in reaching PeV energies. The maximum energy attainable by particles in SNRs can go up to a few hundreds TeV, which is about ten times lower than the energy of the knee. Another class of astrophysical objects considered very promising CR accelerators at PeV energies are young massive stellar clusters (YMSCs) \citep{gabici}.\\
The notion that YMSCs are responsible for the bulk of hadronic CRs at the knee accelerated within our Galaxy represents a viable complement to the long-standing SNR paradigm. Indeed, assuming a termination shock radius of $R_\textrm{ts}\sim10$\,pc around the centre of the YMSC, a magnetic field of $B\sim$10 \textmu G, and a high stellar wind velocity of the order of $\sim$3000 km\,s$^{-1}$ \citep{gabici, morlino} as star cluster physical parameters, according to the Hillas criterion \citep{Hillas, Gab} the maximum energy at which particles can be accelerated is of the order of:
\begin{equation}
E_{\textrm{max}} \sim \frac{v}{3000\,\textrm{km/s}} \,\frac{B}{10\,\textrm{$\mu$G}}\,\frac{R_{\textrm{ts}}}{10\,\textrm{pc}} \sim 1 \,\textrm{PeV}.
\end{equation}
Since CRs are charged particles and are therefore deviated by magnetic fields after escaping their acceleration site, it is in general not possible to reconstruct their trajectory. However, interacting with the gas and the radiation fields that surround the YMSCs, the high-energy CRs produce gamma-rays with energies above 100 TeV, that can be used to identify both hadronic and leptonic PeVatrons. \\
Among the next-generation IACT facilities, the ASTRI Mini-Array (Astrofisica a Specchi con Tecnologia Replicante Italiana) and the CTAO (Cherenkov Telescope Array Observatory) will have a large field of view ($\sim10\,^\circ$ in diameter), an angular resolutions of a few arcmin at 1 TeV, and energy resolutions of $5-10\%$ \citep{ctao, astri}. Both observatories will be capable of detecting radiation up to hundreds of TeV and are expected to greatly improve our knowledge of the morphological properties of extended TeV sources with respect to the instruments currently operating, such as HESS (High Energy Stereoscopic System) or the HAWC (High Altitude Water Cherenkov) experiment. 
\citet{Aharonian1} highlighted that the morphology of YMSCs makes them particularly suited for studies using ground-based Cherenkov telescopes. Their typical angular size of $\sim$$1^\circ$ on the sky allows the CTAO and the ASTRI Mini-Array to resolve spatially these structures, providing valuable insights into CR acceleration driven by stellar winds.\\
\begin{table}
\caption{\label{tab:parametri_ymsc}Main parameters of the YMSCs studied in this work. For each YMSC the columns contain the name, the distance of the object from the Sun (Cygnus OB2: \citealt{2021MNRAS.502.6080O,2019MNRAS.484.1838B}, Westerlund 1: \citealt{Wd1}, Danks 1 \& 2: \citealt{danks}, Markarian 50: \citealt{markarian}), its age (Cygnus OB2: \citealt{Cyg5myr,cygage,cygagelow}, Westerlund 1: \citealt{wd1age,wd1agelow}, Danks 1 \& 2: \citealt{danks}, Markarian 50: \citealt{markarian}), and its luminosity (\citealt{celli}).}
\centering
\begin{tabular}{lccc}
\hline\hline
Cluster Name & Distance (kpc) & Age (Myr) & L$_{\text{kin}}$ (erg s$^{-1}$) \\
\hline
Cygnus OB2      & 1.6   & 1–6    & $2 \times 10^{38}$  \\
Westerlund 1    & 3.9   & 3–10    & $1 \times 10^{39}$  \\
Danks 1         & 3.8   & 1.5    & $8 \times 10^{37}$  \\
Danks 2         & 3.8   & 3      & $7 \times 10^{37}$  \\
Markarian 50    & 3.4   & 7.5   & $9 \times 10^{36}$  \\
\hline
\end{tabular}
\end{table}
This work examines the potential of these telescopes to observe YMSC morphology, resolve their gamma-ray emission, and disinguish them from other extended TeV sources with similar angular sizes, such as TeV halos. For this study, we modelled a sample of five YMSCs, whose key physical parameters are listed in Table \ref{tab:parametri_ymsc}.

\section{The Sample \label{section:sample}}
We based our study on a sample of 5 YMSCs. Since we are interested in studying the emission  up to the $\sim 100$ TeV energy range, we simulate them as if observed by the small-size telescopes (SSTs) present in the most favourable hemisphere: The YMSCs in the northern sky, namely Cygnus OB2 (Cyg OB2), and Markarian 50 (Mrk50), were studied with the ASTRI Mini-Array instrument response function (IRF); the ones in the south, i.e. Westerlund 1 (Wd1), Danks 1 and 2 (Dk1 \& Dk2), with the IRF of the CTAO Southern Array.\\
\begin{table}
\caption{Wind velocities assumed for each YMSC and their computed bubble sizes $R_\text{b}$ and termination shock radii R$_\text{ts}$ in pc and degrees around YMSCs.}
\centering
\begin{tabular}{lcccccc}
\hline\hline
Cluster Name & Wind Speed  & R$_\text{b}$ & R$_\text{ts}$ & R$_\text{b}$ & R$_\text{ts}$ \\
             & (km\,s$^{-1}$) & (pc)& (pc)&($^\circ$) &($^\circ$) \\
\hline
Cygnus OB2   & 2500  & 97.1   & 15.2   & 3.47  & 0.55 \\
Westerlund 1 & 3000  & 159.3  & 25.2   & 2.36  & 0.38 \\
Danks 1      & 3000  & 53.4   & 8.0    & 0.81  & 0.12 \\
Danks 2      & 3000  & 78.8   & 10.1   & 1.19  & 0.15 \\
Markarian 50 & 3000  & 90.5   & 7.9    & 1.53  & 0.13 \\
\hline
\label{tab:rtsrb}
\end{tabular}
\end{table}
The first object in our sample is Cygnus OB2, which is sometimes classified as an open cluster or a star association. We have included it in our sample because of its high mass and luminosity and because of its location in the sky is in spatial correspondence with LHAASO (Large High Altitude Air Shower Observatory) observations of PeV photons \citep{CygLHAASO}. However, the wind velocity is probably lower than the ones in more compact clusters. This is due to the fact that the massive stars are further apart and their winds merge after they have significantly decelerated \citep{vieu_morlino}. In accordance with \citet{Westerlund_hess}, we therefore assumed a 2500 km s$^{-1}$ wind velocity. The distance of this YMSC from the Sun is generally estimated in the 1.6-1.8 kpc range using GAIA data (e.g. \citealp{2021MNRAS.502.6080O} and \citealp{2019MNRAS.484.1838B}), and we chose 1.6 kpc \citep{cyg16}. The age estimations can somewhat vary between 2 and 6 Myr (e.g. \citealp{Cyg5myr}, \citealp{cygage} and \citealp{cygagelow}). \citet{Cyg10myr} find that the star formation activity in Cygnus OB2 has been going on for about 10 Myr, peaking at about 3 Myr ago with the formation of the dense aggregate of very hot and luminous stars that nowadays dominates it. Depending on whether the stellar rotation along their axes is considered and the stellar evolution model assumed, age estimations are affected by 1--2 Myr uncertainties \citep{Cygrot}. The most likely age of Cygnus OB2 of 3--5 Myr, but the absence of SNRs suggests an age $t_{\rm age}\leq3$--$4$ Myr.\\
All the other YMSCs are more compact and therefore believed to produce strong collective winds in the merging of the individual stellar winds \citep{canto_morlino}. In particular, Westerlund 1 is one of the most luminous YMSCs at TeV energies. It has already been studied with HESS data and it has been found to be compatible with the \citet{morlino} model that we therefore employ in the present work \citep{Westerlund_hess, Wd1}. Although we chose a value of 4 Myr as the age of Westerlund 1 (in accordance to \citealp{Wd1_2}), it should be noted that its value has been revisited in recent years and can range from 3 to 10 Myr (e.g. \citealp{wd1age}, \citealp{wd1agelow}). This large uncertainty is due to our limited knowledge of the star formation burst history in the YMSC.\\
The twins Danks 1 and Danks 2 YMSCs were also included in our sample because of their luminosities at TeV energies, which are among the highest according to \citet{mitchell}. They are also in a rather isolated position in the sky and have been potentially detected at GeV energies \citep{Danks_fermi}, making them promising candidates for dedicated observations. We used the age and distance estimations for both Danks 1 and Danks 2 from \citet{danks}.\\
The last YMSC in the sample is Markarian 50, which is located in the northern hemisphere and it is older then the other YMSCs of the sample. Its age suggests that SN explosions could contribute to the total gamma-ray emission, but Markarian 50 is one of the best candidates for observations in the northern sky because of its luminosity \citep{mitchell}. We used the distance and age from \citet{markarian}.\\
We used the physical parameters of Table \ref{tab:parametri_ymsc} to calculate the size of the region of interest around each YMSC, i.e. the bubble excavated in the ISM by the stellar winds from the YMSC itself. We computed also the size of the termination shock radius, at which the bulk of the acceleration is expected to take place. We compute these as respectively \citep{Aharonian1, koo}:
\begin{equation}
\begin{split}
\label{eq:rbrts}
    &R_{\mathrm{b}}\, = 0.88\,L_{\mathrm{w}}^{0.2}\,\rho_{\mathrm{H}}^{-0.2} \,t_{\mathrm{age}}^{0.6}\, \mathrm{pc}\\
    &R_{\mathrm{ts}}= 0.92\,L_{\mathrm{w}}^{0.3} \, v_{\mathrm{w}}^{-0.5} \, \rho_{\mathrm{H}}^{-0.3}\, t_{\mathrm{age}}^{0.4}\,\mathrm{pc}.   
\end{split}
\end{equation}
Here, $L_{\mathrm{w}}$ $v_{\mathrm{w}}$ are the YMSC wind luminosity and velocity in erg\,s$^{-1}$ and km\,s$^{-1}$, respectively, $\rho_{\mathrm{H}}=n_{\mathrm{H}}m_{\mathrm{p}}$ is the gas density around it ($n_{\mathrm{H}}$ is the particle density in cm$^{-3}$ and $m_{\mathrm{p}}$ is the proton mass) and $t_{\mathrm{age}}$ is the age of the YMSC in Myr.\\
A more general version of these equations is the following:\\
\begin{equation}
\begin{split}
\label{eq:eta}
    R_{\mathrm{b}}(t) =& 112 \left( \frac{\eta_m L_{\mathrm{w,c}}}{10^{37}\,\mathrm{erg\,s^{-1}}} \right)^{1/5}
    \left( \frac{n_0}{10\,\mathrm{cm^{-3}}} \right)^{-1/5}
    \left( \frac{t}{10\,\mathrm{Myr}} \right)^{3/5} \, \mathrm{pc}\\
    R_{\mathrm{ts}}(t) =& 26\, \eta_m^{-1/5} 
    \left( \frac{\dot{M}_{\mathrm{c}}}{10^{-4}\,M_{\odot}\,\mathrm{yr^{-1}}} \right)^{3/10}
    \left( \frac{v_{\mathrm{w}}}{2000\,\mathrm{km\,s^{-1}}} \right)^{1/10}\\
    &\left( \frac{n_0}{10\,\mathrm{cm^{-3}}} \right)^{-3/10}
    \left( \frac{t}{10\,\mathrm{Myr}} \right)^{2/5} \, \mathrm{pc},
\end{split}
\end{equation}
where the extra parameter $\eta_m$ is the mechanical efficiency and it represents the fraction of wind energy converted into effective mechanical energy which can inflate the bubble. In the case of an adiabatic scenario, $\eta_m = 1$ and the values of $R_{\text{b}}$ and $R_{\text{ts}}$ revert back to the ones of Equation \ref{eq:rbrts}. If energy losses due to radiative cooling are also considered, however, the efficiency decreases as shown by numerical simulations (e.g. \citealt{eta_morlino}) depending on the age of the cluster and the density of the gas surrounding the YMSC. Following Equation \ref{eq:eta}, lower values of $\eta_m$ imply that the bubble size is smaller and the termination shock radius is larger. For very low efficiencies the termination shock reaches the bubble edge and collapses, therefore we consider only cases of $\eta \ge 0.1$.\\
Following the assumptions of the \citet{morlino} model, we assumed values of wind velocity, mass loss and gas density around the YMSCs of 3000 km s$^{-1}$, $10^{-4}\, M_\odot\,yr^{-1}$ and 10 cm$^{-3}$ respectively, apart from Cygnus OB2, in which case we used a lower wind velocity for the reasoning mentioned above. We also used the wind luminosities computed in \citet{celli}. We show the results of our computations in Table \ref{tab:rtsrb}.\\
In the case of Westerlund 1, we also computed the two radii for different $\eta_m$ values (Table \ref{tab:rbrts_eta}) in order to show how they change as the mechanical efficiency decreases.\\
\begin{table}
\centering
\begin{tabular}{cccccc}
\hline\hline
$\eta_m$  & R$_\text{b}$ & R$_\text{ts}$ & R$_\text{b}$ & R$_\text{ts}$ \\
               & (pc) & (pc) & ($^\circ$) & ($^\circ$) \\
\hline
$0.10$     & 102.4  & 27.4  & 1.54  & 0.41 \\
$0.25$     & 123.0  & 22.8  & 1.85  & 0.34 \\
$0.50$     & 141.3  & 19.9  & 2.13  & 0.30 \\
$0.75$     & 153.3  & 18.3  & 2.31  & 0.28 \\
$1.00$     & 162.4  & 17.3  & 2.45  & 0.26 \\
\hline
\end{tabular}
\caption{Computed sizes of $R_{\mathrm{b}}$ and $R_{\mathrm{ts}}$ around Westerlund 1 for different $\eta_m$ values.}
\label{tab:rbrts_eta}
\end{table}
About $\sim$2200 unidentified sources have been detected by Fermi, and $\sim7\%$ of all Fermi unassociated sources are potentially associated to embedded YMSCs \citep{fermi_morlino}. Furthermore, many TeV and multi-TeV sources remain unidentified \citep{hess_morlino, lhaaso_cat}, among which several YMSCs are expected. One of the goals of our work is to develop methods to identify and classify these sources on a morphological and spectral basis. We therefore compare their morphological features to the ones of TeV halos, which are a class of TeV  sources which presents characteristics similar to YMSCs (Galactic, very large extensions). We wanted to compare their morphological features to the ones of the YMSCs by modelling the secondary gamma-ray emission of both classes of sources. We have considered the TeV halos around the Geminga (PSR J0633+1746) and Monogem (PSR J0659+1414) pulsars \citep{2017Sci...358..911A}. We chose to include TeV halos in our sample of sources because the YMSCs have similar emission characteristics to the halos of Geminga and Monogem, e.g. comparable emission energy ranges and angular extensions of a few degrees.

\section{CTAO and ASTRI Mini-Array Simulations}
We used the software package \texttt{Gammapy} \citep{gammapy:2023, gammapy_zenodo} to compute energy-dependent models of the YMSCs gamma-ray emission. We set the energy range of our simulations to 1--200 TeV. \\
In the case of Geminga and Monogem, our models were obtained by computing the differential photon flux across the TeV halos. We used the \textit{2d} radiation profile presented in \citet{2017Sci...358..911A}, which describes the gamma-ray radiation profiles as a function of the energy $E$ and the angular separation $\theta$ from the centre of the source:
\begin{equation}
    \frac{d^{2}N}{dEd\Omega} = N_{0} \left(\frac{E}{20~\mathrm{TeV}}\right)^{-\alpha} \frac{1.22}{\pi^{3/2} \theta_{d}(E) (\theta + 0.06 \theta_{d}(E))} e^{-\theta^{2}/\theta_{d}(E)^{2}}.
\end{equation}
Here, $\theta_d$ is a diffusion angle proportional to the square root of the diffusion coefficient and is of the order of a few degrees. We generated a model with an angular grid finer than the resolution of the instruments that were used to simulate the observations (i.e. $\sim2.5$ arcmin). The model we used was produced from data in the $8-40$ TeV energy range and we assume that it is valid in the whole energy range we use in our simulations. \\
When producing the YMSC template models, we chose to compute the gamma-ray emission starting from the \citet{morlino} hadronic model, which describes the diffusive shock acceleration of protons at the termination shock radius. Imposing the same angular fineness requirement than the TeV halos on the grids that were the bases of our models, we computed the expected proton fluxes first. Opposed to the case of Geminga and Monogem, the hadronic energy-dependent models of the bubbles around YMSCs are functions of the radial distance from the YMSCs themselves. We therefore computed the proton fluxes in a 3d grid in the same 1--200 TeV energy range. With the Naima libraries \citep{naima}, we converted the proton model to a secondary gamma-ray model assuming pion decay as the main photon production mechanism. Finally, we integrated along the line of sight to have a \texttt{Gammapy}-compatible differential flux template model.\\
We have simulated the observation of the YMSCs using the entire ASTRI Mini-Array (i.e. 9 telescopes) IRF \cite{astriirf} and the 20$^\circ$ zenith southern site CTAO IRF \citep{ctairf}. The results presented in this paper were all obtained with a 200\,hr observational live-time and with the sources simulated on-axis. We simulated all sources individually, except for Danks 1 and Danks 2 due to their proximity in the sky (0.05$^\circ$). For each of the YMSCs, and we considered the non-astrophysical background that includes the CR background, the diffuse atmospheric background, the instrumental background, and the sky background relative to the selected field of view (FoV) through the appropriate \texttt{Gammapy} libraries. The emission from the two TeV halos and their backgrounds were simulated using the ASTRI Mini-Array IRF, While both Geminga and Monogem are observable in the winter months from the Southern hemisphere with elevation angles up to 40.5$^\circ$ and 44.0$^\circ$ respectively, we study them as they can be observed from the most favourable hemisphere (i.e. the Northern one) since they reach higher elevation angles of $79.4^\circ$ and $75.9^\circ$ for Geminga and Monogem respectively. \\
We simulated observations of an $8\times8$ deg$^2$ FoV, so that the angular dimensions of all the YMSCs could fit within it.

\section{Source Classification\label{sec:source_class}}
\begin{figure}
\centering
\includegraphics[width=0.75\linewidth]{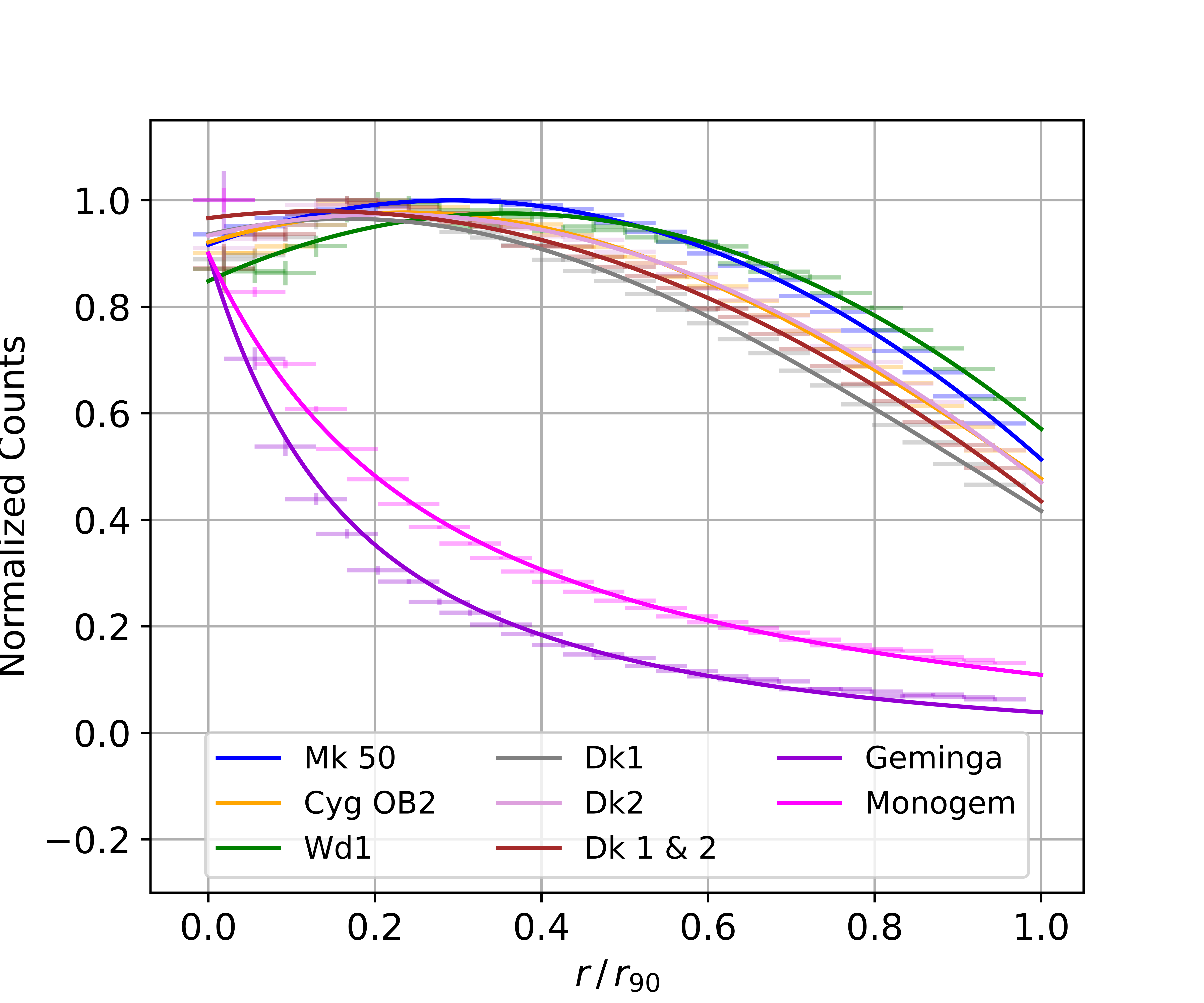}
\caption{Gamma-ray emission radial profiles as a function of the distance from the centre for all YMSCs and TeV halos in the sample. The profiles have been rescaled by dividing by the maximum of their radial emission for clarity of comparison. Here the data are fitted with the better fitting function for each object (i.e. the modified Gaussian or the modified Lorentzian), according to the relative $\chi^2$ values in Table \ref{tab:chi2}.}
\label{fig:radial_profiles_spheric}
\end{figure}
\begin{figure}
\centering
\includegraphics[width=0.75\linewidth]{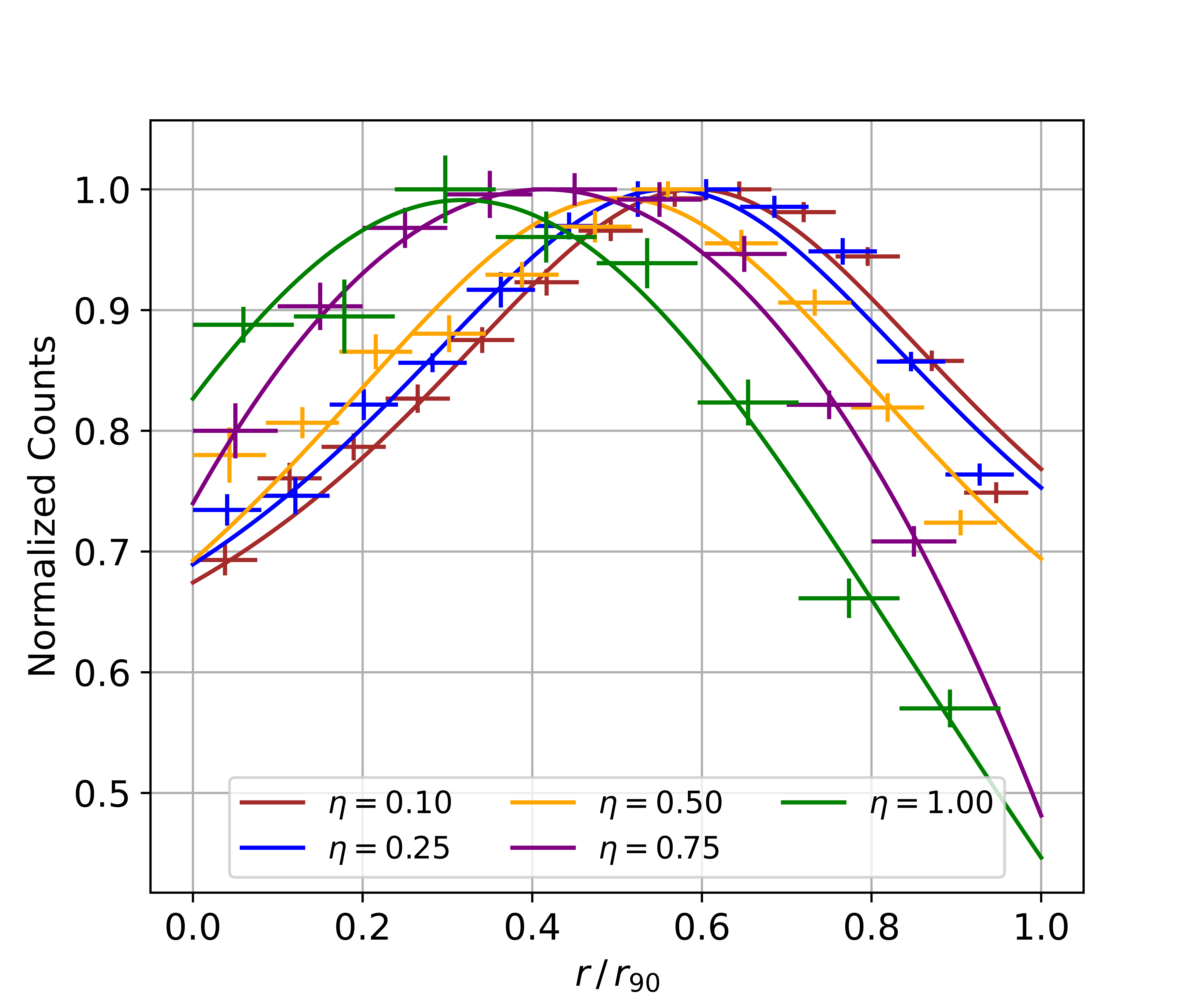}
\caption{Westerlund 1 gamma-ray emission radial profiles as a function of the distance from the centre for varying values of $\eta_m$. The profiles have been rescaled as before and the data are fitted with the modified Lorentzian function.}
\label{fig:profiles_eta}
\end{figure}
\begin{figure}
\centering
\includegraphics[width=0.75\linewidth]{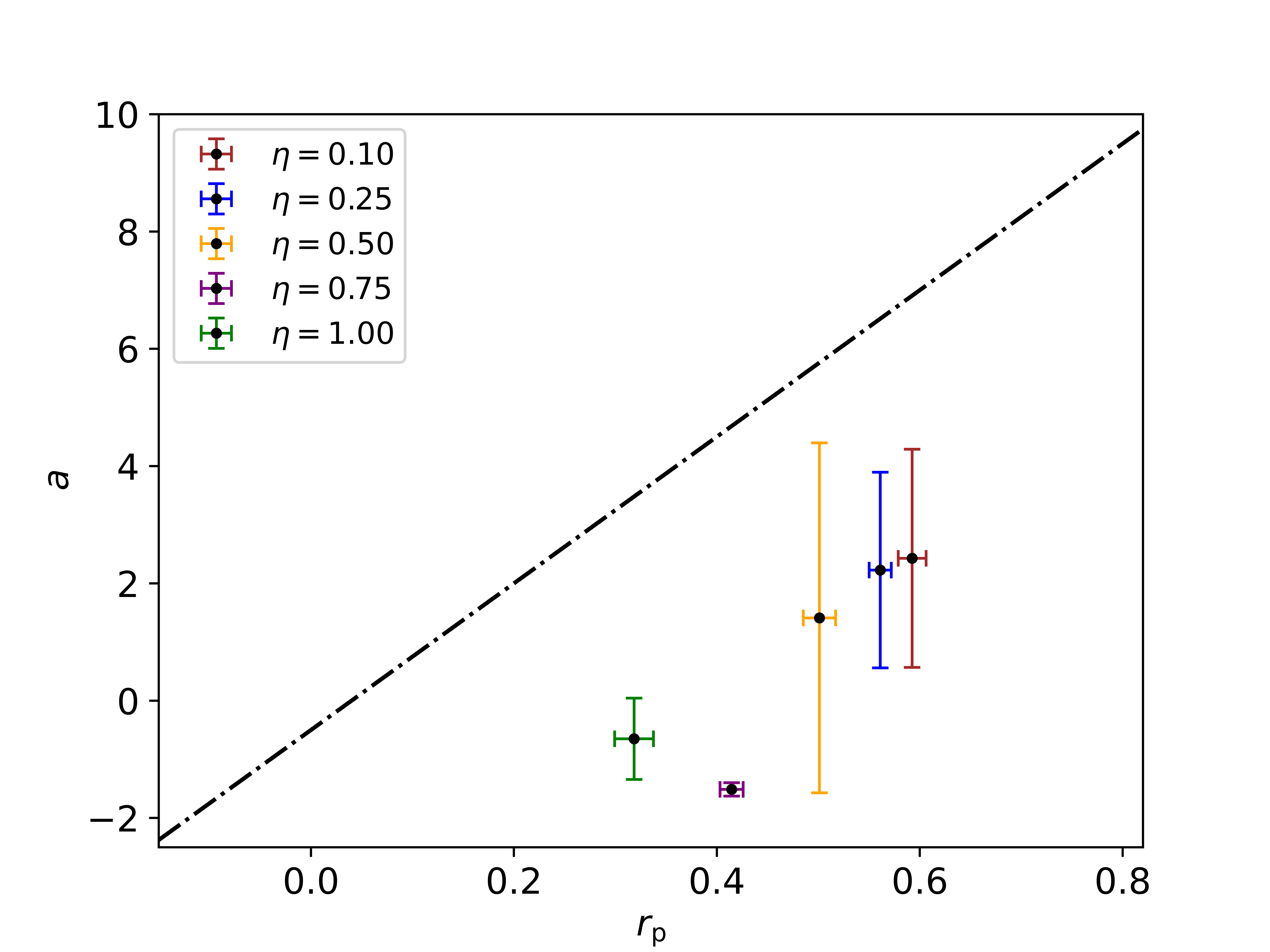}
\caption{Values of the curve anisotropy and emission peak position in units of $r_{90}$ obtained with the modified Lorentzian fit to the data. For all $\eta_m$ values, Westerlund 1 displays $a$ and $r_{p}$ values in a region below the dot-dashed line.}
\label{fig:rpa_eta}
\end{figure}

In order to study the morphology of the sources as they will be observed, we inspected the gamma-ray emission radial profiles, i.e. the gamma-ray emission as a function of the distance from the centre of sources. In particular, we were interested in modeling the excess radiation maps (the background-subtracted gamma-ray counts map) by parameterizing and comparing the behaviour of the YMSCs and the TeV halos. We opted for computing the excess counts per deg$^2$ by correcting the emission in each bin of the profiles for the area of the respective ring around the centre of the source. We also corrected for the effective area in each pixel of the simulated map and we estimated the observed position of the objects in two different ways depending on the nature of the simulated source: we computed the peak of the gamma-ray emission in the case of TeV halos and the barycentre of the emission for YMSCs, given the spherical symmetry of the model. We then used the source position as the centre of the rings and estimated the observed radius of both the halos and the YMSC bubble by measuring the 90\% containment radius of their excess counts, i.e. the radius of a circular region within which 90\% of detected photons is contained. We also re-scaled the amplitude of the profiles by dividing them by their respective maximum so that they are more easily comparable.\\
By summing the emission in the whole simulated 1--200 TeV energy range, we obtained the radial profiles shown in Figure \ref{fig:radial_profiles_spheric}. We rescaled them over $r_{90}$ so that $0<r<1$ and we tried the following two functions for the fit:
\begin{equation}
    f_\textrm{g}(r;  r_\textrm{p}, N, a, w) = N \, \exp\left\{-\left[\frac{(1 + e^{a \, (r-  r_\textrm{p}))} }{w} \, (r-  r_\textrm{p})\right]^2\right\},
\end{equation}
\begin{equation}
    f_\textrm{l}(r;  r_\textrm{p}, N, a, w) = \frac{N}{1 + \left(\frac{r-  r_\textrm{p}}{w}\right)^2} \, \left[1 + a \, (r-  r_\textrm{p})^2\right].
    \label{mod_lor}
\end{equation}
\begin{figure}
\centering
\includegraphics[width=0.75\linewidth]{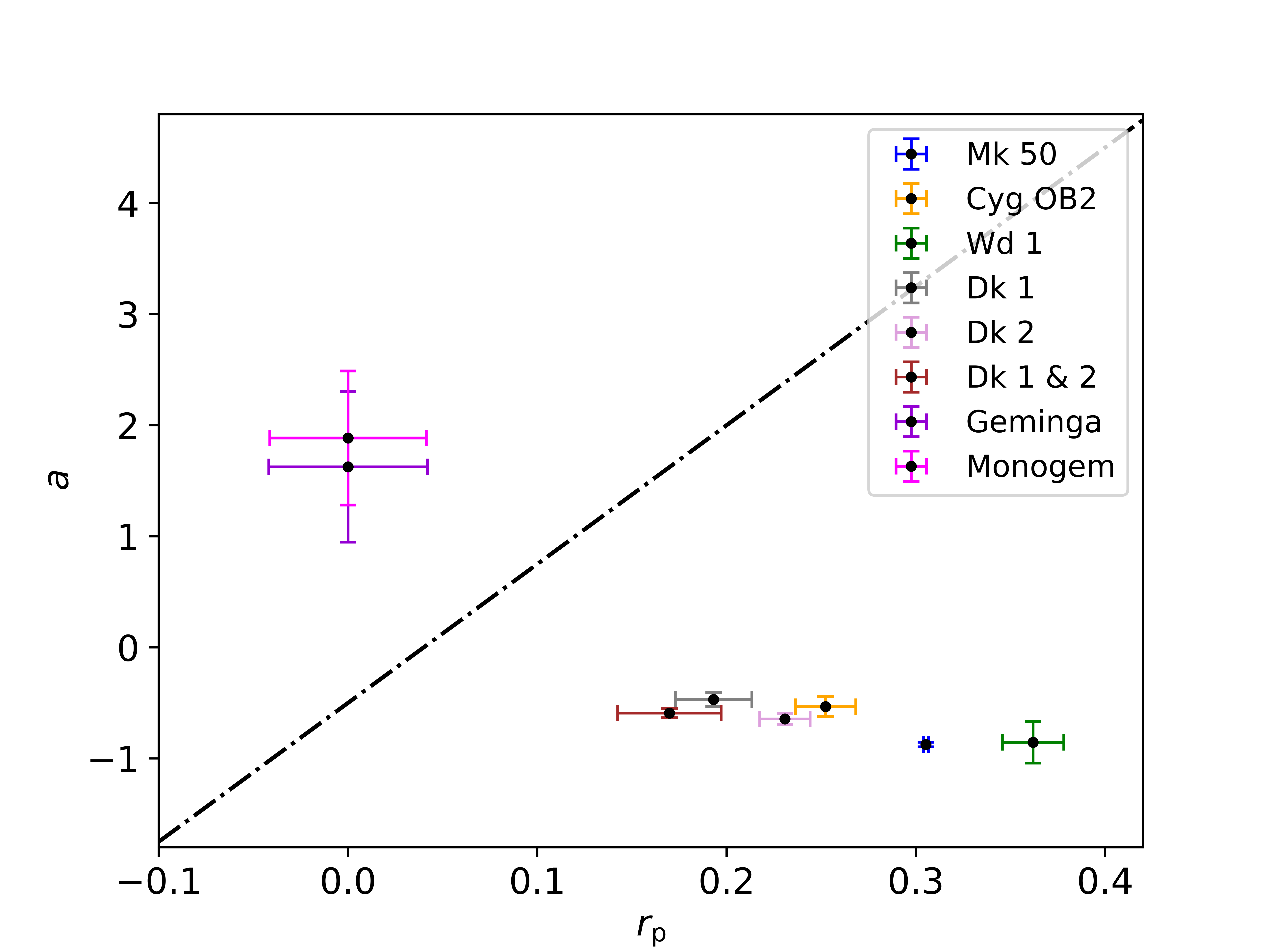}
\caption{Values of the curve anisotropy and emission peak position in units of $r_{90}$ obtained with the modified Lorentzian fit to the data. In all the cases studied, YMSCs display lower $a$ values and higher $r_{p}$ values, positioning themselves in the lower-right corner. On the other hand, TeV halos are on the opposite corner of the plot, in a region ideally above the dot-dashed line.}
\label{fig:r_a_scatter_plot}
\end{figure}
The first one is a Gaussian function modified to account for the anisotropy of the profile peaks; the second is similar, but it is based on a Lorentzian function. The Lorentzian function is intended to account for higher tails that can be especially relevant in modelling the emission profiles of extended sources. In both functions, $N$ is the normalization, while $w$ and $ r_\text{p}$ are the curve width and the peak position in units of the 90\% confinement radius. The $a$ parameter quantifies the anisotropy of the curve with respect to its peak.\\
\begin{sidewaystable}
\caption{\label{tab:chi2}“Fit parameters for the simulated data using both fitting functions” (transposed). The data have been binned into 50 intervals between the centre of each source and its 90\% containment radius so that the bin width roughly corresponds to the angular resolution of the ASTRI Mini-Array and the CTAO. The reduced $\chi^2$ have been calculated over the number of degrees of freedom $n_\textrm{dof} = 46$.}
\centering
\begin{tabular}{lcccccc}
\hline\hline
Source & $\chi^2_\textrm{g}$ & $r_\textrm{p,g}/r_{90}$ & $a_\textrm{g}$ & $\chi^2_\textrm{l}$ & $r_\textrm{p,l}/r_{90}$ & $a_\textrm{l}$ \\
\hline
Cygnus OB2      & 1.837  & $0.11 \pm 0.07$ & $1.0 \pm 0.1$   & 1.155  & $0.25 \pm 0.02$ & $-0.98 \pm 0.03$ \\
Westerlund 1    & 0.279  & $0.2 \pm 0.1$   & $1.9 \pm 0.2$   & 0.823  & $0.37 \pm 0.01$ & $-1.28 \pm 0.09$ \\
Danks 1         & 1.219  & $0.10 \pm 0.06$ & $0.7 \pm 0.1$   & 0.813  & $0.20 \pm 0.03$ & $-0.81 \pm 0.03$ \\
Danks 2         & 0.808  & $0.10 \pm 0.07$ & $1.1 \pm 0.1$   & 0.990  & $0.23 \pm 0.01$ & $-0.95 \pm 0.03$ \\
Danks 1 \& 2    & 1.365  & $0.10 \pm 0.06$ & $1.0 \pm 0.1$   & 1.132  & $0.19 \pm 0.02$ & $-0.93 \pm 0.02$ \\
Markarian 50    & 7.251  & $0.282 \pm 0.005$ & $0.80 \pm 0.05$ & 11.424 & $0.312 \pm 0.004$ & $-0.99 \pm 0.04$ \\
Geminga         & 5.961  & $0.00 \pm 0.08$ & $-3.1 \pm 3.1$  & 0.902  & $0.00 \pm 0.05$ & $1.7 \pm 0.6$ \\
Monogem         & 54.932 & $0.01 \pm 0.04$ & $-2.4 \pm 2.5$  & 2.631  & $0.00 \pm 0.04$ & $2.6 \pm 0.5$ \\
\hline
\end{tabular}
\end{sidewaystable}
The complete results of the fits can be found in Table \ref{tab:chi2}. Both functions are fitted to the simulated data, as shown by the reduced $\chi^2$ values. Though the use of the modified Lorentzian leads to better $\chi^2$ values for the radial profiles of both YMSCs and TeV halos, the modified Gaussian function results in a better fit in the cases of the Markarian 50 YMSC. The profiles of the YMSCs show a behaviour characterised by an emission peak at larger distances from the centre of the systems with respect to TeV halos.\\
Of the parameters resulting from our fits, we find that the anisotropy parameter $a$, together with emission peak position $ r_{p}$, characterise very well the two classes of objects that we have studied. As it can be seen in Figure \ref{fig:r_a_scatter_plot} for the case of the modified Lorentzian function, the combination of the two parameters can be a powerful tool to classify an unidentified source as either a YMSC or a TeV halo using its morphology. While the emission peak position can somewhat vary depending on the physical parameters of the YMSCs, we find that the peak anisotropy is a reliable parameter to characterise the morphology and evaluate the nature of a source. The peak anisotropy and position $ r_\textrm{p}$ parameters obtained with the fit of the modified Gaussian function yields similar conclusions, but $a$ is less constrained. Still, YMSCs are characterised by smaller values of $a$ and higher values of $ r_\textrm{p}$.\\
Taking the case of Westerlund 1 as an example, we also show how values of $\eta_m\le1$ impact our analyses. We simulated observations similar to the ones described above, but we computed the proton and gamma-ray spectra assuming the different values of $\eta_m$ from Table \ref{tab:rbrts_eta}. As shown in Figure \ref{fig:profiles_eta}, we used the modified Lorentzian function as a fit to the data and confirmed that the $r_p$ and $a$ parameters are still within acceptable bounds for our classification criteria. Figure \ref{fig:rpa_eta} shows that the ($r_p$,$a$) points are below the dot-dashed line, as expected for YMSCs. In particular, the $r_p$ parameter is remarkably useful because its value is higher for lower mechanical efficiencies, since the $R_{\text{ts}}/R_\text{b}$ ratio increases. We therefore expect our classification methods to hold for $\eta_m\le1$, and to even work better for higher $R_{\text{ts}}/R_\text{b}$ ratios.\\
\begin{figure}
\centering
\includegraphics[width=0.75\linewidth]{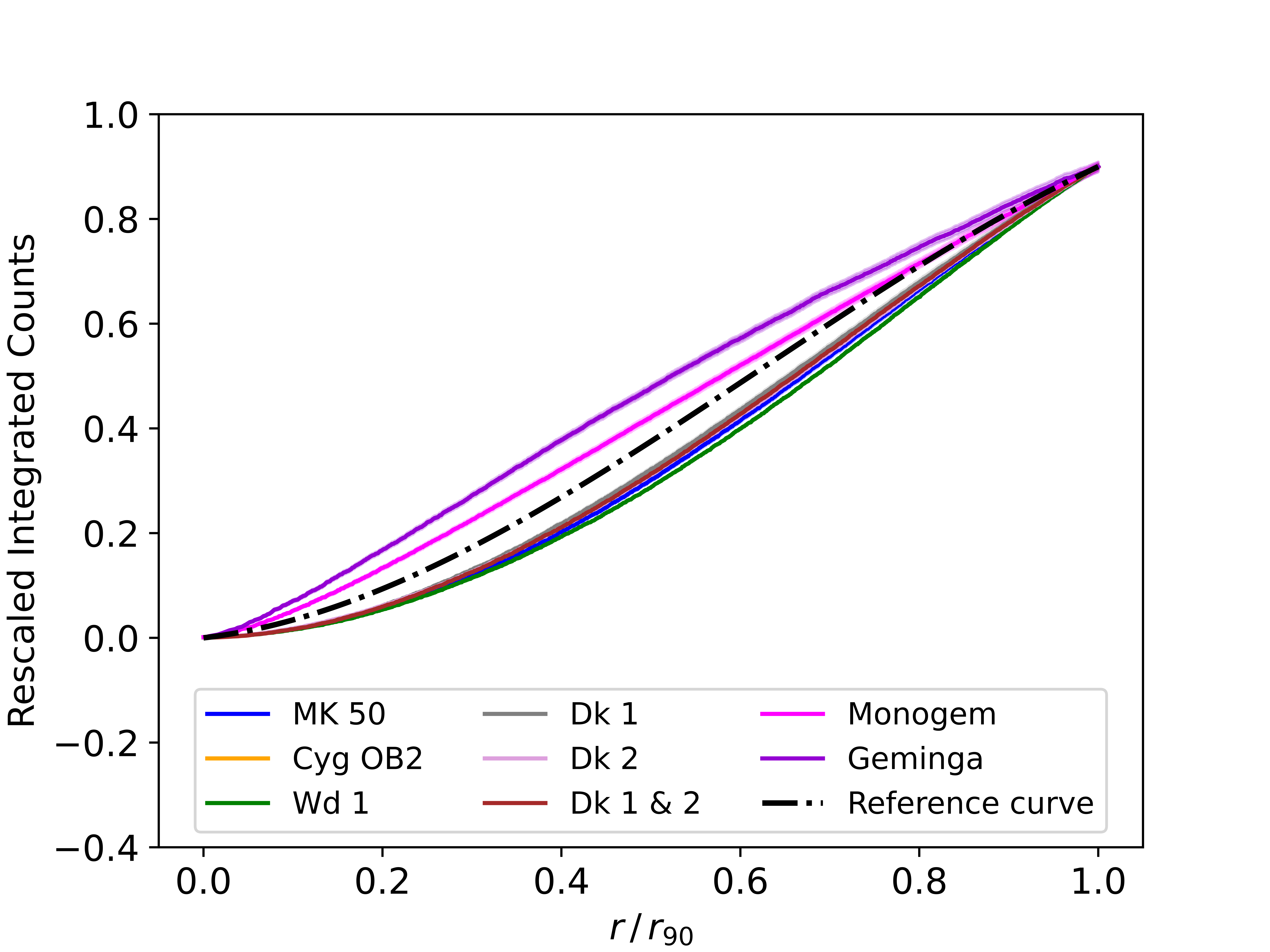}
\caption{Integrated counts profile curves of the emission from the centres of each source. The curves have been re-scaled over the total emission for clarity of comparison. Both YMSCs and TeV halos are shown, with the reference function (black dot-dashed curve) of Equation \ref{eq:int} dividing the two classes of objects: TeV halos are above the reference curve and YMSCs are below.}
\label{fig:cumulatice_profiles_spheric}
\end{figure}
The integrated radial excess profiles can also provide valuable morphological information and they can be used as an alternative mean of classification of unidentified YMSCs. Figure \ref{fig:cumulatice_profiles_spheric} shows the cumulative counts of both each YMSCs and TeV halos. For clarity of comparison, the above re-scaling was also used here. The integrated profiles of Geminga and Monogem present a very different behaviour compared to YMSCs: counts increase at a smaller distance from the centre. This implies that the profiles of the two  classes of objects lay in different sectors of the plane and we chose a reference curve (dot-dashed line in Figure \ref{fig:cumulatice_profiles_spheric}) that produces two plane sectors:
\begin{equation}
    f_\textrm{r}(r) = -0.8\,r^3 + 1.5\,r^2 +0.2\,r.
    \label{eq:int}
\end{equation}
The area differences between the reference curve and the integrated profile of each source \textit{i}:
\begin{equation}
A_i =\int^1_0   f_\textrm{r}(r)\, dr \,\,- \,\, \int^1_0 f_\textrm{i}(r)\, dr  
\end{equation}
are listed in Table \ref{tab:object_areas} and serve as an indication of the confidence in source classification. In our case, all sources are correctly classified, with positive values for all YMSCs and negative values for the Geminga and Monogem TeV halos.
\begin{table}
\caption{\label{tab:object_areas}Areas between the reference curve and the integrated profiles for the considered sources. Positive areas indicate that the profile curve is below the reference function $f_\textrm{r}(r)$, while negative areas indicate it is above.}
\centering
\begin{tabular}{lc}
\hline\hline
Object         & Area \\
\hline
Mk 50          & 0.045 \\
Cyg OB2        & 0.040 \\
Wd 1           & 0.054 \\
Dk 1           & 0.034 \\
Dk 2           & 0.040 \\
Dk 1 \& 2      & 0.039 \\
Geminga        & -0.062 \\
Monogem        & -0.027 \\
\hline
\end{tabular}
\end{table}

\section{LHAASO Sources Simulations and Analyses}

\begin{figure}[t]
    \centering
    \begin{subfigure}{0.75\linewidth}
        \centering
        \includegraphics[width=\linewidth]{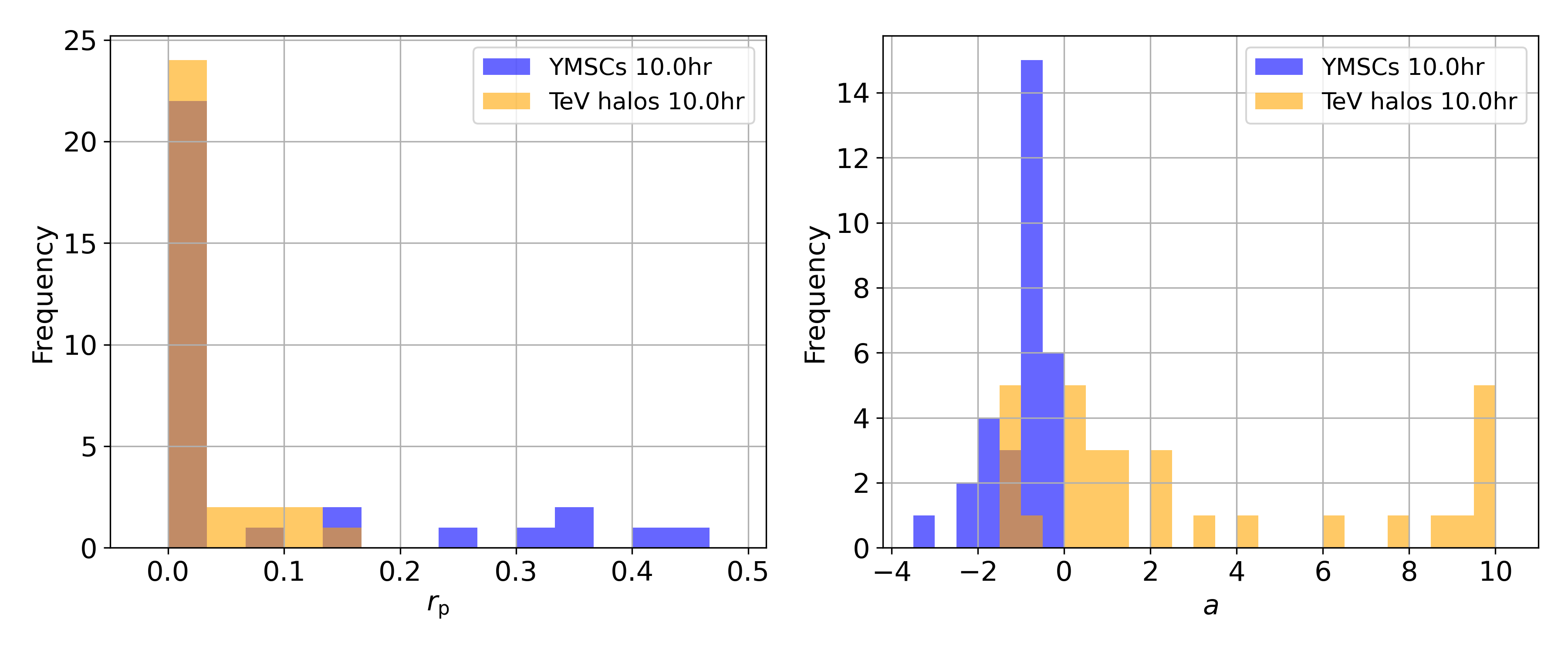}
        \caption{10\,hr simulations distributions.}
    \end{subfigure}\\
    \begin{subfigure}{0.75\linewidth}
        \centering
        \includegraphics[width=\linewidth]{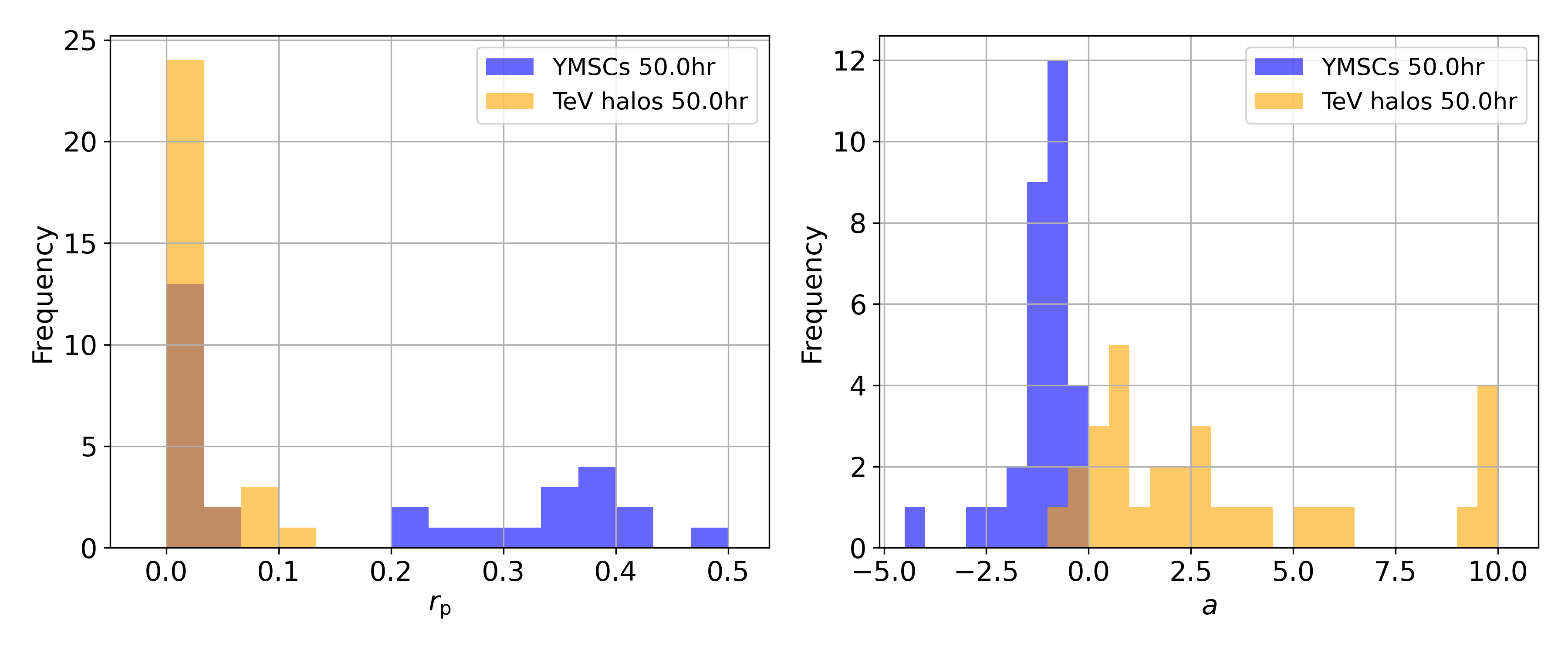}
        \caption{50\,hr simulations distributions.}
    \end{subfigure}\\
    \begin{subfigure}{0.75\linewidth}
        \centering
        \includegraphics[width=\linewidth]{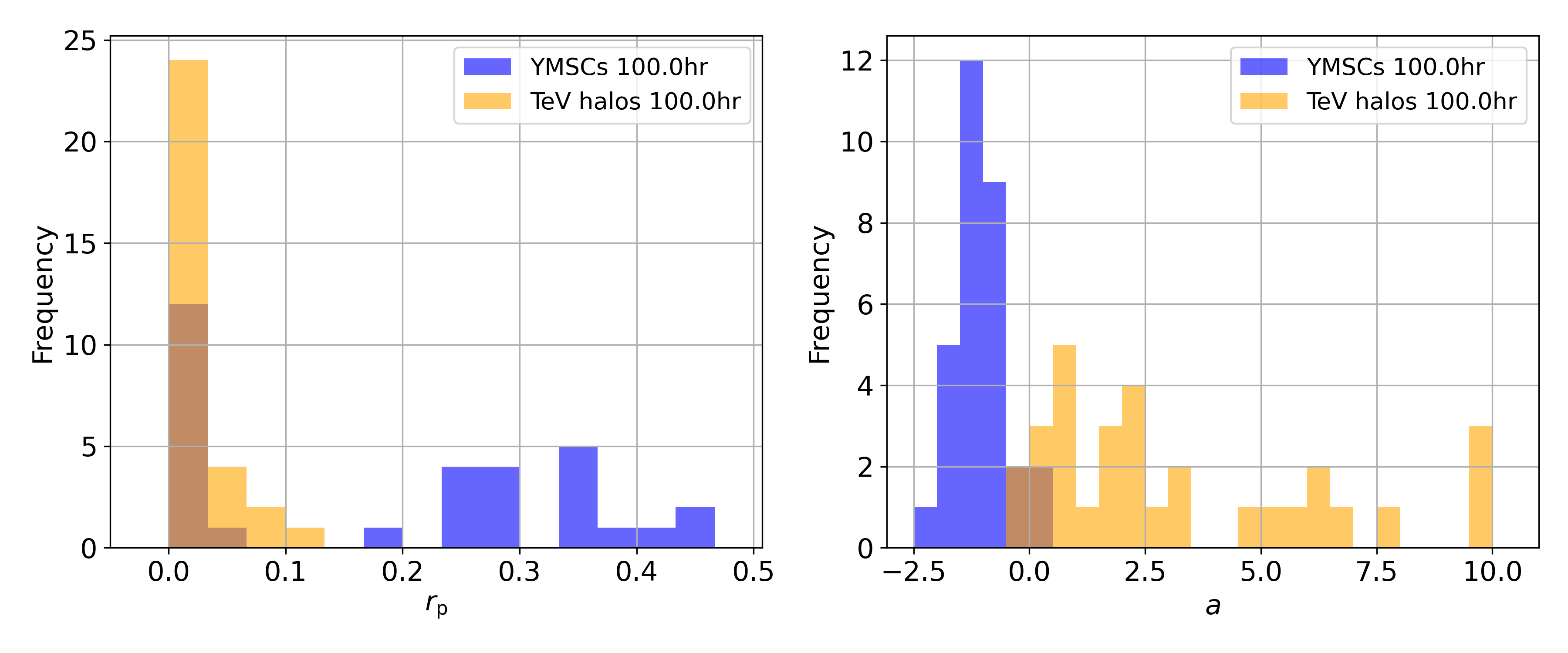}
        \caption{100\,hr simulations distributions.}
    \end{subfigure}
    
    \caption{Peak position (\textit{left}) and anisotropy (\textit{right}) parameters distributions for the listed livetimes. Blue distributions are the results of YMSC simulations with the CTAO South IRF, orange ones are derived from TeV halos simulations. Brown counts indicate that the estimated peak position of the YMSC and TeV halo simulations coincide.}
    \label{fig:histogram_cta}
\end{figure}

\begin{figure}[t]
    \ContinuedFloat
    \centering
    \begin{subfigure}{0.75\linewidth}
        \centering
        \includegraphics[width=\linewidth]{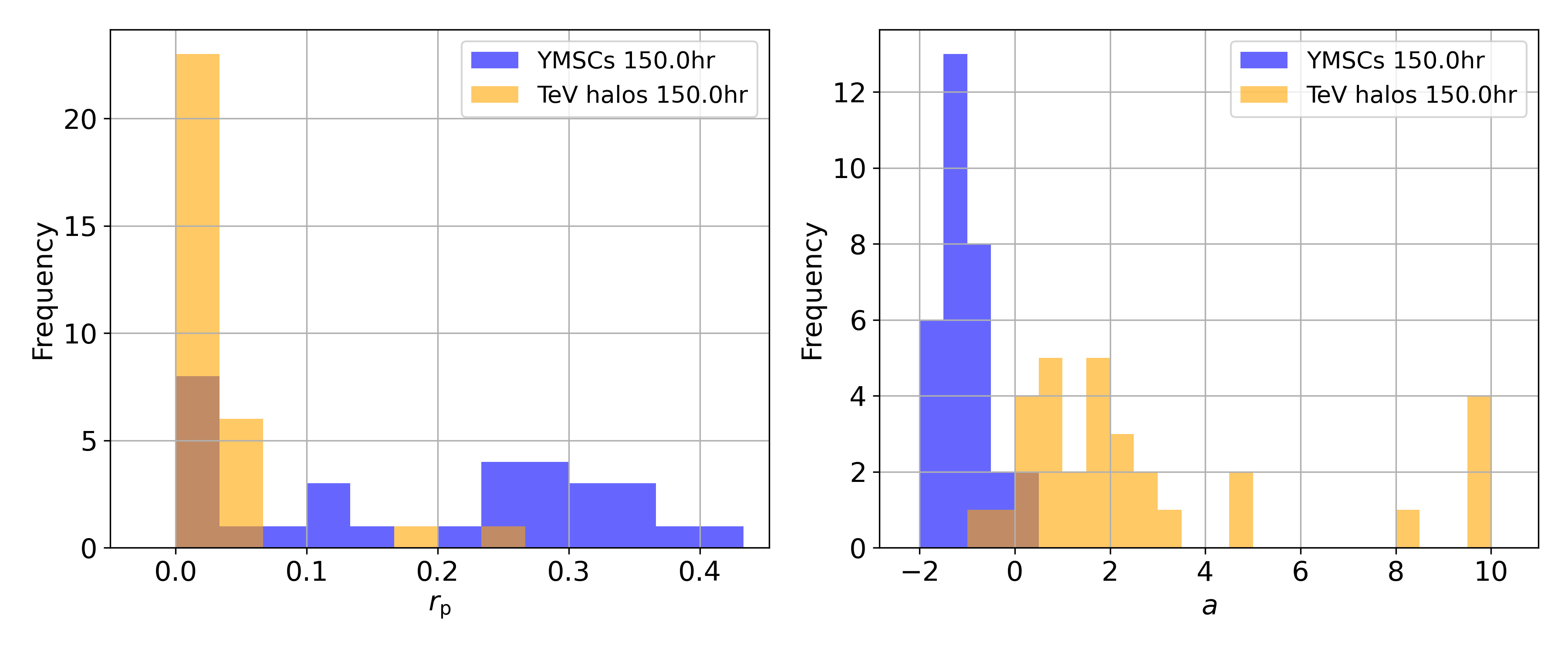}
        \caption{150\,hr simulations distributions.}
    \end{subfigure}\\
    \begin{subfigure}{0.75\linewidth}
        \centering
        \includegraphics[width=\linewidth]{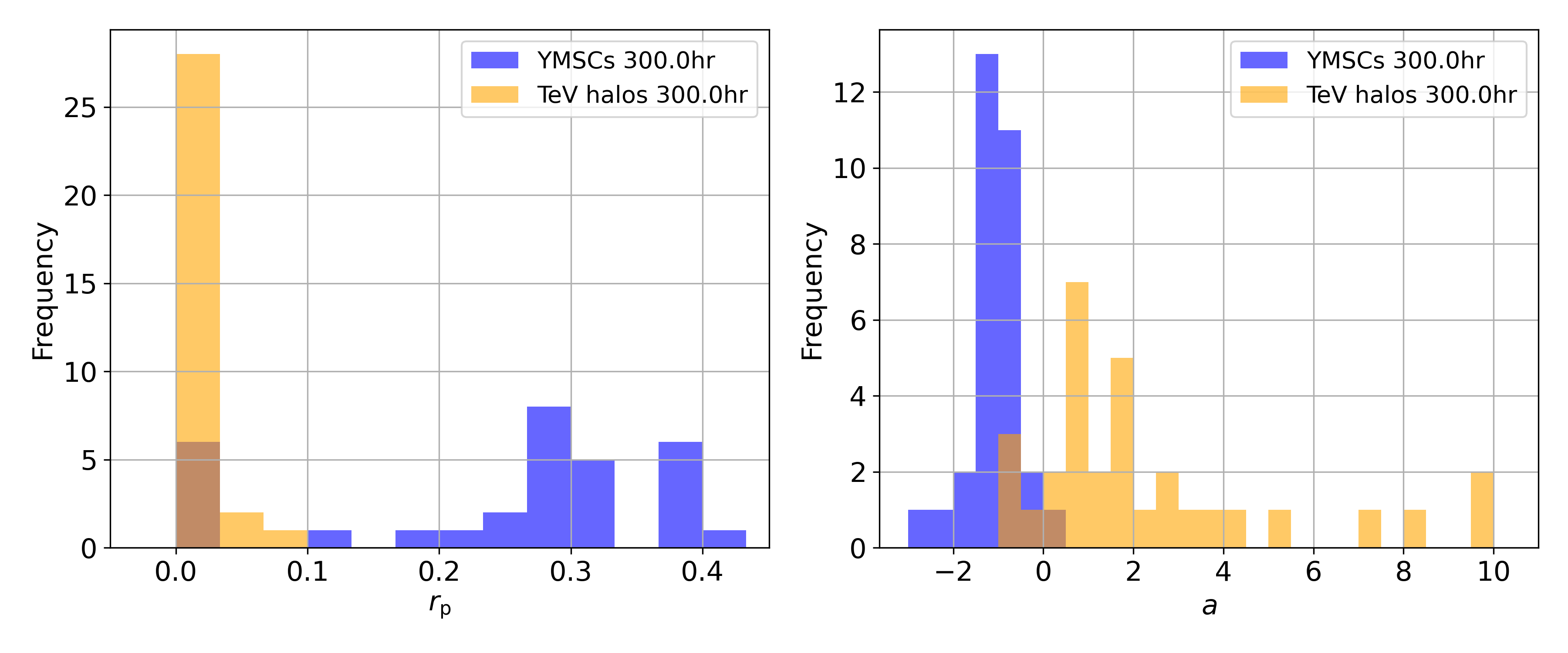}
        \caption{300\,hr simulations distributions.}
    \end{subfigure}

    \caption{Continuation.}
\end{figure}

\begin{figure}[t]
    \centering
    \begin{subfigure}{0.75\linewidth}
        \centering
        \includegraphics[width=\linewidth]{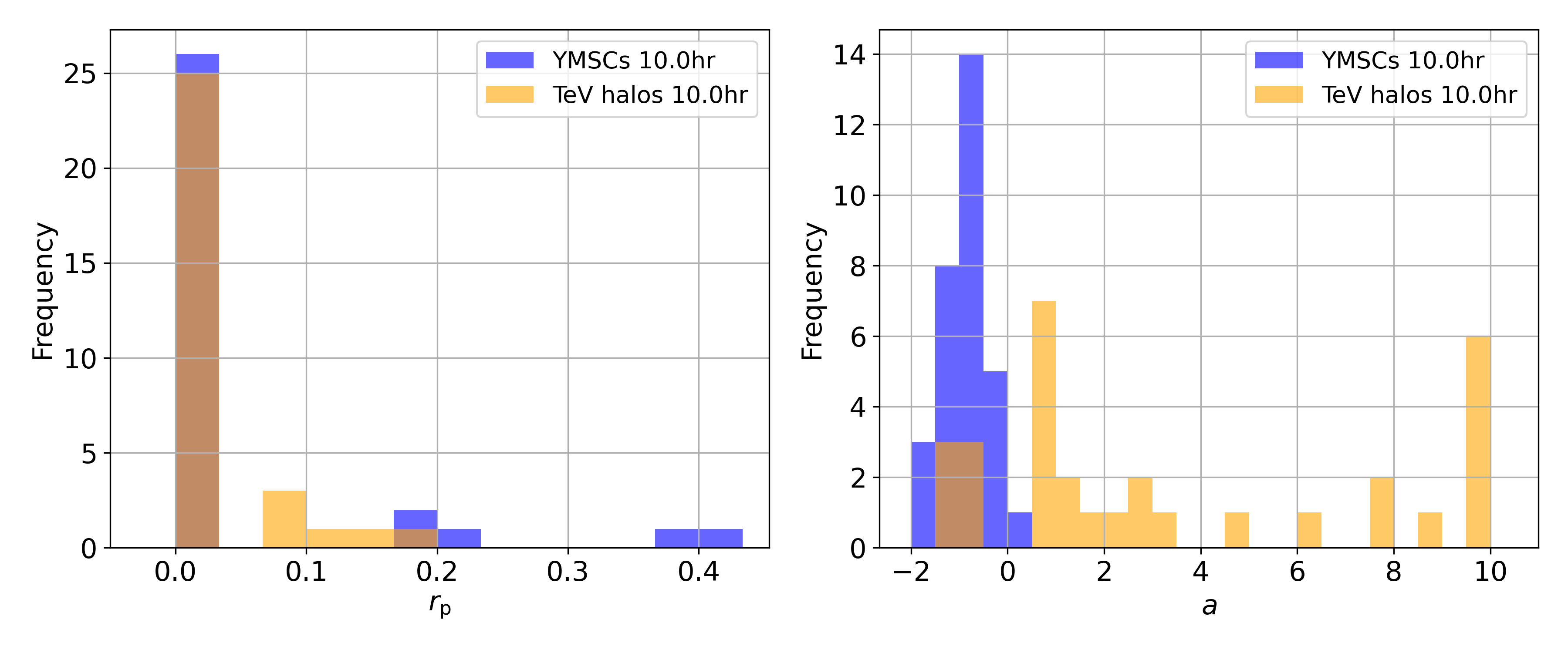}
        \caption{10\,hr simulations distributions.}
    \end{subfigure}
    
    \caption{Same as Figure~\ref{fig:histogram_cta}, but for simulations with the ASTRI Mini-Array IRF.}
    \label{fig:histogram_astri}
\end{figure}

\begin{figure}[t]
    \ContinuedFloat
    \centering
    \begin{subfigure}{0.75\linewidth}
        \centering
        \includegraphics[width=\linewidth]{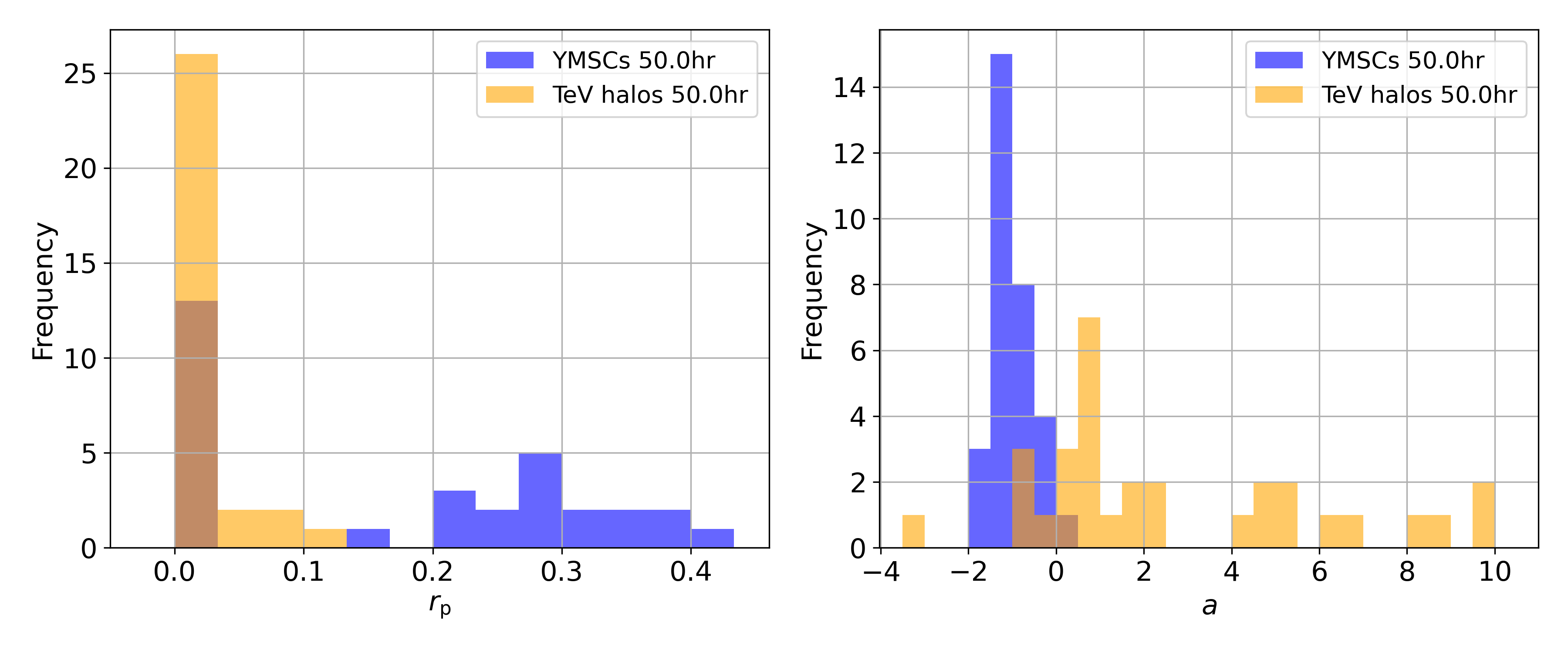}
        \caption{50\,hr simulations distributions.}
    \end{subfigure}\\
    \begin{subfigure}{0.75\linewidth}
        \centering
        \includegraphics[width=\linewidth]{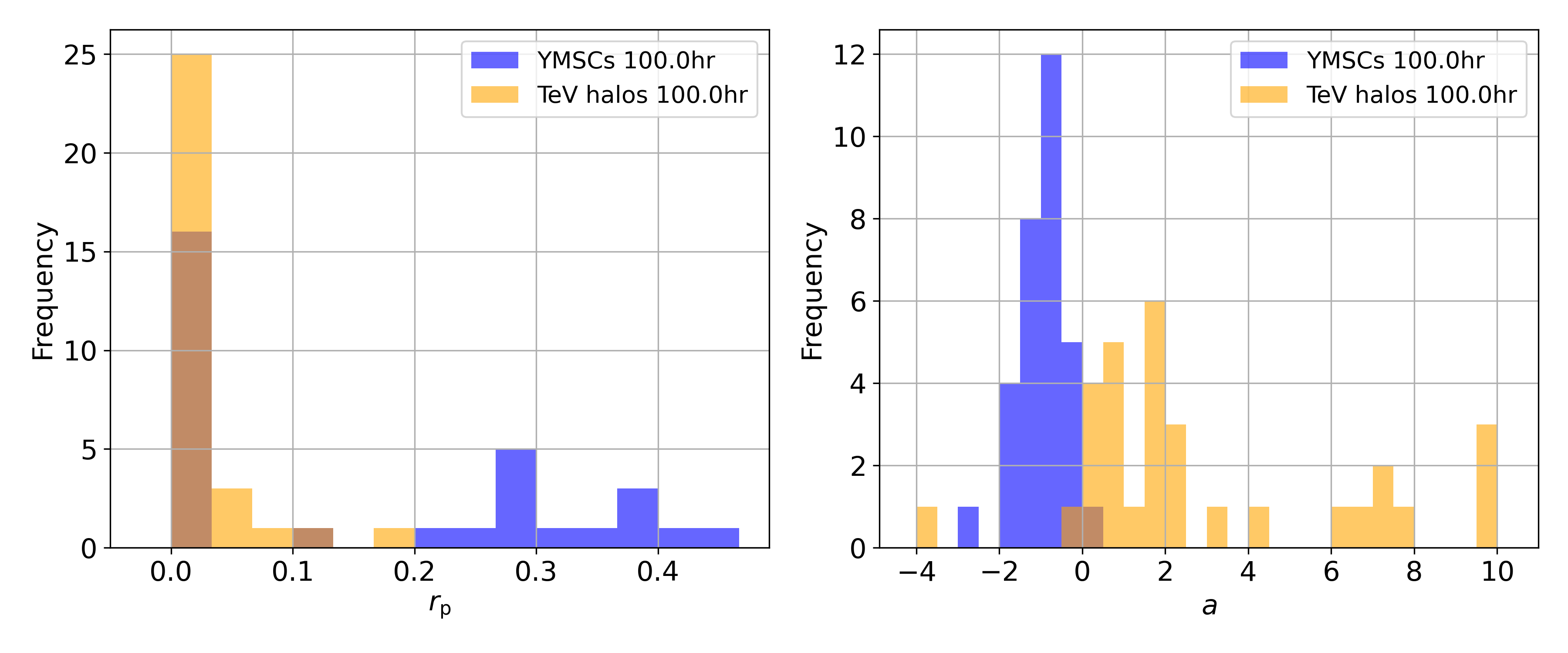}
        \caption{100\,hr simulations distributions.}
    \end{subfigure}\\
    \begin{subfigure}{0.75\linewidth}
        \centering
        \includegraphics[width=\linewidth]{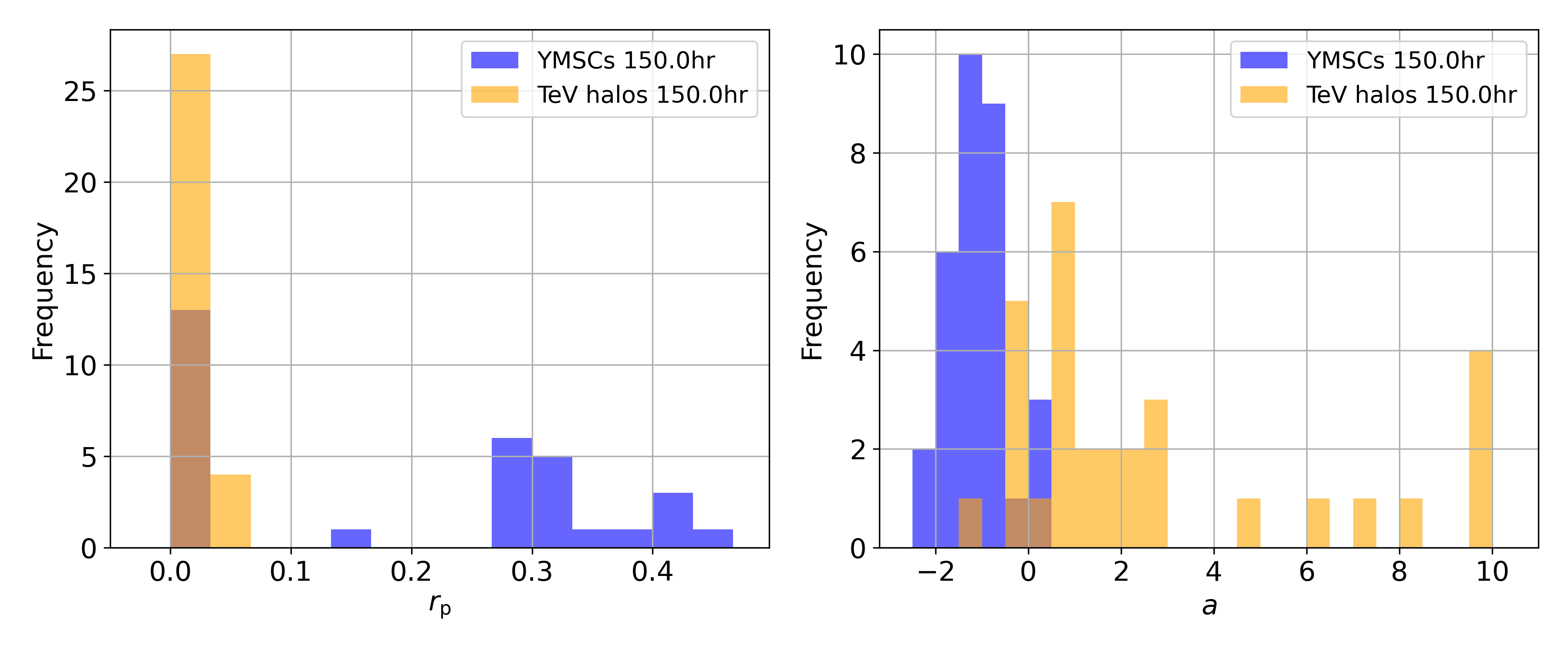}
        \caption{150\,hr simulations distributions.}
    \end{subfigure}\\
    \begin{subfigure}{0.75\linewidth}
        \centering
        \includegraphics[width=\linewidth]{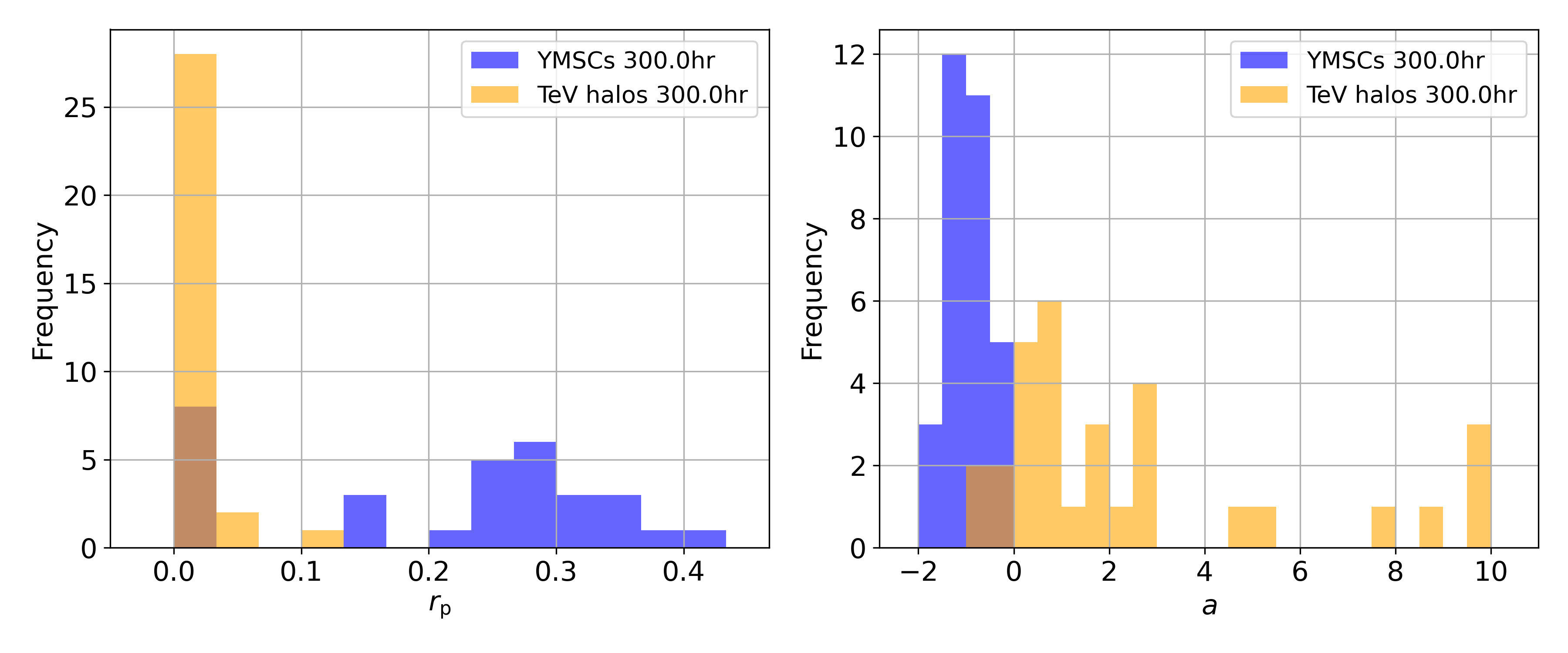}
        \caption{300\,hr simulations distributions.}
    \end{subfigure}

    \caption{Continuation.}
\end{figure}

\begin{sidewaystable}
\caption{\label{tab:histogram}Percentage of correct classifications, misclassifications, and inconclusive classification according to the fit $r_p$ and $a$ parameters from simulations performed using the CTAO South and the ASTRI Mini-Array IRFs for different livetimes. The parameters are grouped by source type: YMSCs and TeV halos.}
\centering
\begin{tabular}{l|ccc|ccc}
\hline\hline
\multicolumn{7}{c|}{\textbf{CTAO South}}\\
\hline
 & \multicolumn{3}{c|}{\textbf{YMSC Sources}} & \multicolumn{3}{c}{\textbf{TeV Halo Sources}} \\
\cline{2-7}
Livetime & Classified & Misclassified & Inconclusive & Classified & Misclassified & Inconclusive \\
         & (\%) & (\%) & (\%) & (\%) & (\%) & (\%) \\
\hline
10 h     & 12.9 & 3.2 & 83.9 & 6.5 & 6.5 & 87.0 \\
25 h     & 19.4 & 3.2 & 77.4 & 22.6 & 3.2 & 74.2 \\
50 h     & 48.4 & 0.0 & 51.6 & 35.5 & 0.0 & 64.5 \\
75 h     & 51.6 & 6.5 & 41.9 & 32.3 & 0.0 & 67.7 \\
100 h    & 54.8 & 0.0 & 45.2 & 48.4 & 0.0 & 51.6 \\
150 h    & 58.1 & 3.2 & 38.7 & 64.5 & 3.2 & 32.3 \\
300 h    & 74.2 & 0.0 & 25.8 & 77.4 & 0.0 & 22.6 \\
\hline
\end{tabular}

\vspace*{1cm}

\begin{tabular}{l|ccc|ccc}
\hline\hline
\multicolumn{7}{c|}{\textbf{ASTRI Mini-Array}}\\
\hline
 & \multicolumn{3}{c|}{\textbf{YMSC Sources}} & \multicolumn{3}{c}{\textbf{TeV Halo Sources}} \\
\cline{2-7}
Livetime & Classified & Misclassified & Inconclusive & Classified & Misclassified & Inconclusive \\
         & (\%) & (\%) & (\%) & (\%) & (\%) & (\%) \\
\hline
10 h     & 6.5 & 3.2 & 90.3 & 3.2 & 0.0 & 96.8 \\
25 h     & 16.1 & 0.0 & 83.9 & 19.4 & 3.2 & 77.4 \\
50 h     & 38.7 & 0.0 & 61.3 & 38.7 & 0.0 & 61.3 \\
75 h     & 45.2 & 3.2 & 51.6 & 32.3 & 3.2 & 64.5 \\
100 h    & 51.6 & 0.0 & 48.4 & 41.9 & 0.0 & 58.1 \\
150 h    & 54.8 & 0.0 & 45.2 & 58.1 & 0.0 & 41.9 \\
300 h    & 64.5 & 0.0 & 35.5 & 71.0 & 0.0 & 29.0 \\
\hline
\end{tabular}
\end{sidewaystable}

In order to test the classification capabilities of the CTAO and the ASTRI Mini-Array, we applied our study to the unidentified extended sources of the first LHAASO catalog \citep{lhaaso_cat}. We selected all non-point sources detected with either or both LHAASO main instruments with an angular extension of $3\sigma<4^\circ$, to make sure that they would fit well within a single-pointing observation. This resulted in a sample of 29 and 26 sources for the Water Cherenkov Detector Array (WCDA) and the km$^2$ Array (KM2A), respectively. Of those, 22 were detected by both LHAASO instruments. Sources detected with both the WCDA and the KM2A are reported in the LHAASO catalog with two independent source models, each composed of a spectral component (i.e. a power-law spectrum with normalization $N_0$ and index $\gamma$) and a spatial component (i.e. a 2d Gaussian shape).\\
\begin{figure}
\centering
\includegraphics[width=0.75\linewidth]{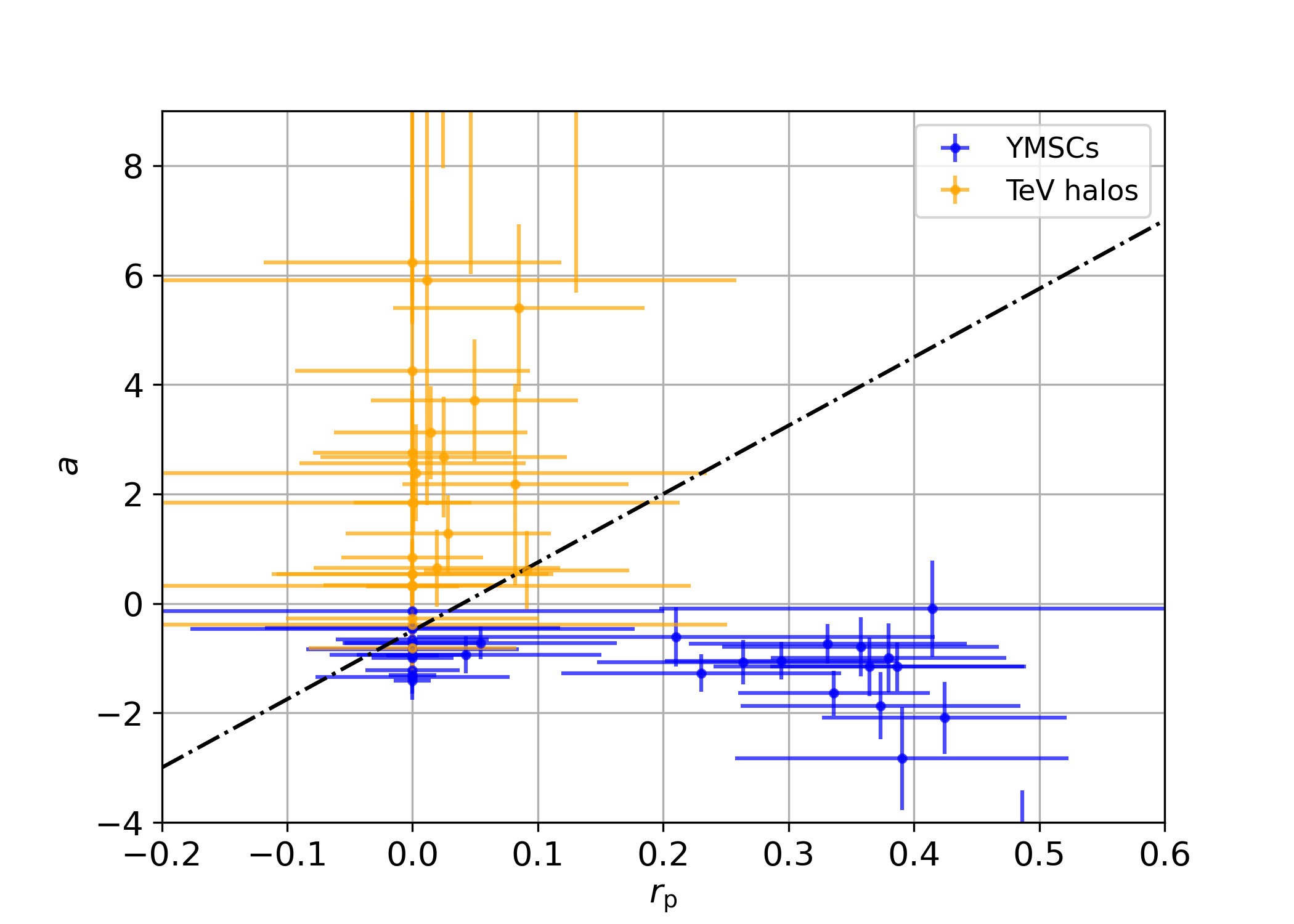}
\caption{Values of the curve anisotropy and emission peak position in units of $r_{90}$ of all the LHAASO sources simulated with the CTAO South IRF. Orange crosses are used for TeV halos and blue crosses are used for YMSCs. The dot-dashed, black line is the same from Figure \ref{fig:r_a_scatter_plot}.}
\label{fig:scatter_cta}
\end{figure}
\begin{figure}
\centering
\includegraphics[width=0.75\linewidth]{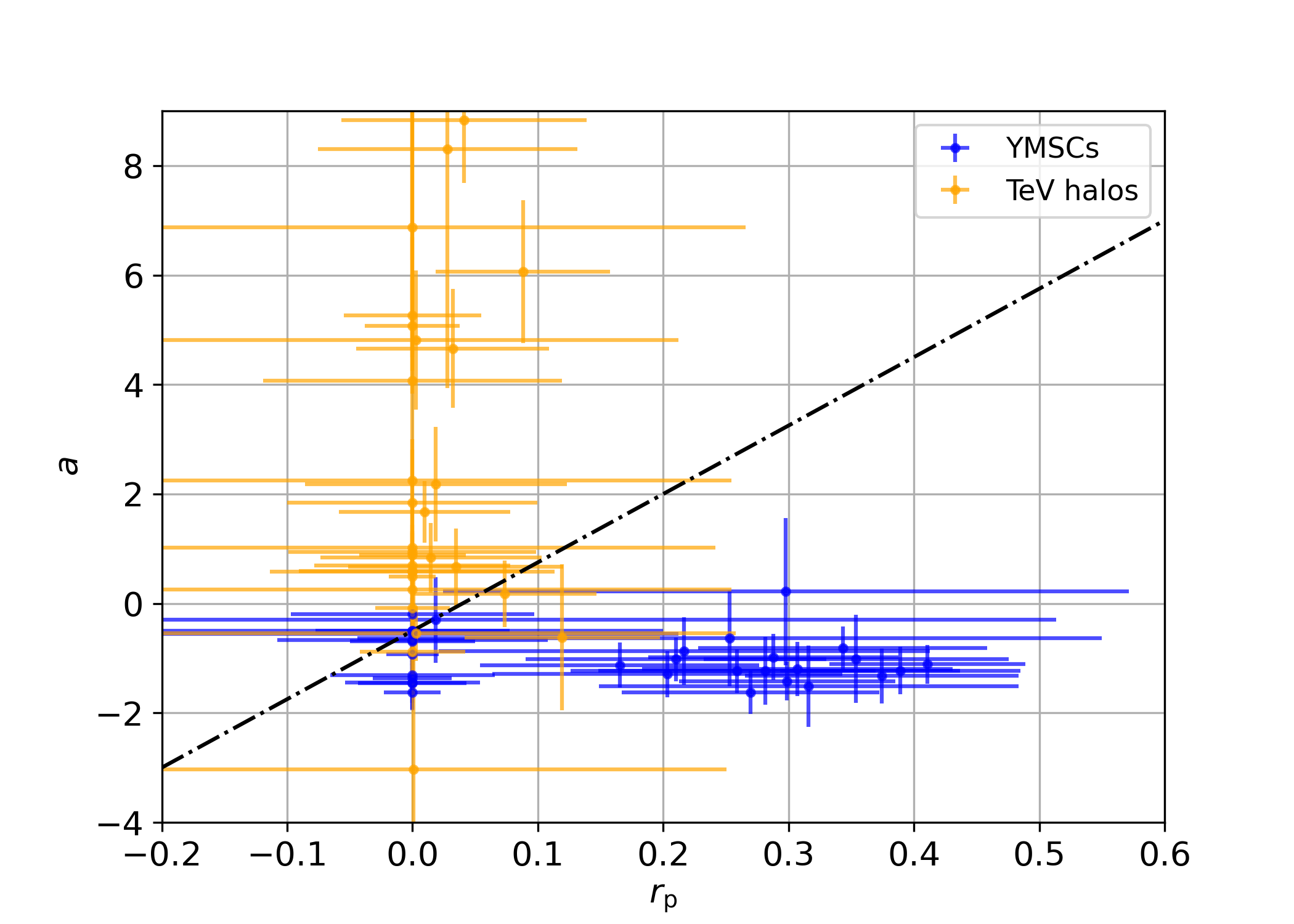}
\caption{Same as Figure \ref{fig:scatter_cta}, but for the case of simulations with the ASTRI Mini-Array IRF.}
\label{fig:scatter_astri}
\end{figure}
We created the correspondent 3d models needed to perform our \texttt{Gammapy} simulations of the sources that met the spatial model criteria (i.e. 2d Gaussian-shaped sources with a $0<\sigma<1.33^\circ$). We assumed a morphology constant in energy and produced two models with the same LHAASO spectrum and angular extension for each source. In the case of a WCDA and KM2A joint detection, we modelled the components of the emission in their respective energy ranges using the relative source angular extensions $\sigma_{WCDA}$ and $\sigma_{KM2A}$. As spectral model for sources detected by both LHAASO instruments we used a broken power-law joint at 25\,TeV. The spectral indexes, normalizations and reference energies were taken from the LHAASO catalog \citep{lhaaso_cat}. In the case of sources detected by either only the WCDA or the KM2A, we used a simple power law with the relative LHAASO parameters. As a spatial profile, we used template models with a spatial extension of $3\sigma$ and a radial profile shape following the one of YMSCs, and then the one of TeV halos deduced in Section \ref{sec:source_class}.\\
We simulated observations with both the CTAO South and the ASTRI Mini-Array IRFs of the 62 resulting source models, where the models of WCDA and KM2A were combined for the mutually detected sources according to their energy ranges, i.e. 1--25 TeV and 25--300 TeV respectively. We generated independent observations centred in the position of the source reported by the LHAASO instrument. In the case of a detection with both WCDA and KM2A, we centred the observation pointing on the average coordinates weighted on the number of expected excess photons. In the latter case, the centre of the observations coincided with the actual centre of the emission due to the higher number of photons at lower energies. 
We once more computed the radial profiles corrected for instrumental effects given by each simulation and repeated the process for different observation livetimes (300 hr, 150 hr, 100 hr, 75 hr, 50 hr, 25 hr and 10 hr) in order to find the minimum amount of hours needed to  assess reliably the nature of the LHAASO unidentified extended sources considered. We then analysed the radial profiles by fitting the data using the modified Lorentzian function in Equation \ref{mod_lor} to estimate the new peak position and anisotropy. \\
The distribution of the peak position and the anisotropy parameters we found in our fits are summarised in Table \ref{tab:histogram} 
and can be seen in Figures \ref{fig:histogram_cta} and \ref{fig:histogram_astri} for the CTAO South and ASTRI Mini-Array IRFs, respectively, and for all the livetimes mentioned above. We find that the emission peak position is clearly distinguishable between the YMSC and TeV halo models (i.e. the estimated $\Delta r_\text{p} \ge 0.15r_\text{90}$) for all livetimes in the case of 6 sources when analysing data with the CTAO South IRF and 3 sources when using the ASTRI Mini-Array IRF. This indicates that these sources can be classified correctly with only 10 hours of observation. In the case of the ASTRI Mini-Array, the successful identification of the emission peak position rapidly increases to 38.7\% and 46.8\% of the total number of sources at 50 and 100\,hr, respectively. With the CTAO South IRF, classification performances are slightly better: 42.0\%  and 51.6\% are correctly identified at 50 and 100\,hr. Prolonged observations of 300\,hr should allow the correct classification of the 67.8\% and 75.8\%  of sources on the basis of the emission peak position using the CTAO South and ASTRI Mini-Array IRFs, respectively. However, a better classification can be operated taking into account both the emission peak and the anisotropy parameter, as explained in Section \ref{sec:source_class}. This can also be seen in Figures \ref{fig:scatter_cta} and \ref{fig:scatter_astri}, where the $a-r_\text{p}$ plots for the 50\,hr livetime simulation results are shown as an example for both the ASTRI Mini-Array and CTAO South cases. In the first case, combining the $r_\text{p}$ and the $a$ values we get only 8 misclassifications out of the 62 simulations, i.e. fits that yield points in the opposite sector with respect to the expected one. Furthermore, 5 of those can be considered as uncertain classifications due to the error boxes of the points intersecting the dot-dashed line. Using the latter IRF, only 6 fits lead to incorrect classifications, all of whom are uncertain according to the same criterion as before. In total, 68.2\% and 65.2\% of sources can be correctly classified with 50\,hr of dedicated observations, combining the information of the two fit parameters.

\section{Conclusions}
In this work, we presented new criteria for the classification of extended source based on their gamma-ray morphology at $E\geq1$ TeV. We based our study on a sample of 5 YMSCs and 2 TeV halos and studied how they will be observed with two of the next-generation IACT telescopes, CTAO South and the ASTRI Mini-Array.\\
We modeled the emission by characterizing the radial excess profile of each source. We found two functions that fit the simulated data with good accuracy and we identified the combination between the emission peak position and the anisotropy of the curves as the main signatures of the nature of the source. 
We also showed that the integrated profiles provide more useful information for the source classification and we proposed a reference function that divides the YMSCs sector from the TeV halos one. We also verified that our classification methods hold even assuming a YMSCs mechanical efficiency $\eta_m<1$.\\
We tested our classification methods on the 7-sources sample that we have considered and we found that they are effective in all cases. In fact, all sources were correctly classified using the curve anisotropy around the peak and the emission peak position parameters. Furthermore, the integrated profiles of each source lays in the expected sector of the plane with respect to the reference curve.\\ 
Overall, we established viable and effective methods to distinguish between YMSCs and TeV halos based only on their morphology at TeV energies. This can potentially improve the classification of unidentified extended sources that the next generation of IACT telescopes will detect, using their improved angular resolution. In this work, we applied our classification methodology to the unidentified extended sources reported in the first LHAASO catalog. By simulating their observations with the CTAO South and the ASTRI Mini-Array IRFs, we evaluated the feasibility of distinguishing their nature based on their morphology. Our results show that, depending on the observation livetime, a significant fraction of the considered sources can be reliably classified. By combining both the emission peak and anisotropy parameters, we achieve a higher accuracy in source identification, with up to 68.2\% and 65.2\% of sources correctly classified in the case of 50 hours of dedicated observations for the CTAO South and the ASTRI Mini-Array, respectively. The sources that remain challenging to classify are typically characterized by either a very low flux per steradian, a small angular extension, or a combination of both, which limits the ability to extract clear morphological signatures. \\
When new ASTRI Mini-Array and CTAO South data in the very-high-energy domain will be available, our classification approach will help in identify extended sources detected with wide-field instruments like LHAASO, contributing to a more comprehensive understanding of their nature and underlying physical processes and to a better understanding of the particle acceleration mechanisms around YMSCs.\\
We plan on addressing several aspects in future works, which are expected to further improve the robustness and applicability of our classification methodology. First, while the present study focuses on YMSCs and TeV halos, other extended sources such as SNRs, PWNe, or composite systems may display similar radial profiles or angular sizes, particularly when distance effects are considered. A systematic comparison with existing catalogues will allow us to verify whether the morphological signatures used here remain discriminating across a broader population of sources and to identify potential degeneracies in classification. Secondly, the energy dependence of source morphology will be studied further. Finally, the impact of observational parameters such as zenith angle warrants further investigation. Zenith angle affects the effective energy threshold, point-spread function, and background rates of IACT instruments, which in turn can influence the quality of the reconstructed morphology. Studying the performance of CTAO South, CTAO North and the ASTRI Mini-Array under different zenith angles will help to determine the range of conditions under which our classification methods remain effective and will support a more direct comparison between different instruments.

\section*{Acknowledgements} 
This publication was produced while A.B. was attending the PhD program in in Space Science and Technology at the University of Trento, Cycle XXXVIII, with the support of a scholarship financed by the Ministerial Decree no. 351 of 9th April 2022, based on the NRRP - funded by the European Union - NextGenerationEU - Mission 4 "Education and Research", Component 1 "Enhancement of the offer of educational services: from nurseries to universities” - Investment 4.1 “Extension of the number of research doctorates and innovative doctorates for public administration and cultural heritage” - CUP C53C22000430006.\\
A.B. acknowledges financial support from the European Union—Next Generation EU under the project IR0000012—CTA+ (CUP C53C22000430006), announcement N.3264 on 28/12/2021: “Rafforzamento e creazione di IR nell’ambito del Piano Nazionale di Ripresa e Resilienza (PNRR)”.\\
M.R acknowledges the NRRP - funded by the European Union - NextGenerationEU (CUP C53C22000430006). This research has made use of the CTA instrument response functions provided by the CTA Consortium and Observatory, see https://www.ctao-observatory.org/science/cta-performance/ (version prod5 v0.1; \cite{ctairf}) for more details.\\
This research has made use of the ASTRI Mini-Array Instrument Response Functions (IRFs) provided by the ASTRI Project \cite{astriirf}.\\
This work made use of Gammapy \citep{gammapy:2023}, a community-developed Python package. The Gammapy team acknowledges all Gammapy past and current contributors, as well as all contributors of the main Gammapy dependency libraries: \hyperref[https://numpy.org/]{NumPy}, \hyperref[https://scipy.org/]{SciPy}, \hyperref[http://www.astropy.org]{Astropy}, \hyperref[https://astropy-regions.readthedocs.io/]{Astropy Regions}, \hyperref[https://scikit-hep.org/iminuit/]{iminuit}, \hyperref[https://matplotlib.org/]{Matplotlib}.



\bibliographystyle{elsarticle-harv} 
\bibliography{references_new}

\begin{thebibliography}{40}
\expandafter\ifx\csname natexlab\endcsname\relax\def\natexlab#1{#1}\fi
\providecommand{\url}[1]{\texttt{#1}}
\providecommand{\href}[2]{#2}
\providecommand{\path}[1]{#1}
\providecommand{\DOIprefix}{doi:}
\providecommand{\ArXivprefix}{arXiv:}
\providecommand{\URLprefix}{URL: }
\providecommand{\Pubmedprefix}{pmid:}
\providecommand{\doi}[1]{\href{http://dx.doi.org/#1}{\path{#1}}}
\providecommand{\Pubmed}[1]{\href{pmid:#1}{\path{#1}}}
\providecommand{\bibinfo}[2]{#2}
\ifx\xfnm\relax \def\xfnm[#1]{\unskip,\space#1}\fi
\bibitem[{{Abdollahi} et~al.(2022){Abdollahi}, {Acero} and {Baldini et al.}}]{fermi_morlino}
\bibinfo{author}{{Abdollahi}, S.}, \bibinfo{author}{{Acero}, F.}, \bibinfo{author}{{Baldini et al.}, L.}, \bibinfo{year}{2022}.
\newblock \bibinfo{title}{{Incremental Fermi Large Area Telescope Fourth Source Catalog}}.
\newblock \bibinfo{journal}{\apjs} \bibinfo{volume}{260}, \bibinfo{pages}{53}.
\newblock \DOIprefix\doi{10.3847/1538-4365/ac6751}, \href{http://arxiv.org/abs/2201.11184}{{\tt arXiv:2201.11184}}.
\bibitem[{{Abeysekara} et~al.(2017){Abeysekara}, {Albert} and {Alfaro et al.}}]{2017Sci...358..911A}
\bibinfo{author}{{Abeysekara}, A.U.}, \bibinfo{author}{{Albert}, A.}, \bibinfo{author}{{Alfaro et al.}, R.}, \bibinfo{year}{2017}.
\newblock \bibinfo{title}{{Extended gamma-ray sources around pulsars constrain the origin of the positron flux at Earth}}.
\newblock \bibinfo{journal}{Science} \bibinfo{volume}{358}, \bibinfo{pages}{911--914}.
\newblock \DOIprefix\doi{10.1126/science.aan4880}, \href{http://arxiv.org/abs/1711.06223}{{\tt arXiv:1711.06223}}.
\bibitem[{Acero et~al.(2025)Acero, Aguasca-Cabot and Bernete~et al.}]{gammapy_zenodo}
\bibinfo{author}{Acero, F.}, \bibinfo{author}{Aguasca-Cabot, A.}, \bibinfo{author}{Bernete~et al., J.}, \bibinfo{year}{2025}.
\newblock \bibinfo{title}{Gammapy: Python toolbox for gamma-ray astronomy}.
\newblock \URLprefix \url{https://doi.org/10.5281/zenodo.14760974}, \DOIprefix\doi{10.5281/zenodo.14760974}.
\bibitem[{{Aharonian} et~al.(2022){Aharonian}, {Ashkar} and {Backes et al.}}]{Westerlund_hess}
\bibinfo{author}{{Aharonian}, F.}, \bibinfo{author}{{Ashkar}, H.}, \bibinfo{author}{{Backes et al.}, M.}, \bibinfo{year}{2022}.
\newblock \bibinfo{title}{{A deep spectromorphological study of the {\ensuremath{\gamma}}-ray emission surrounding the young massive stellar cluster Westerlund 1}}.
\newblock \bibinfo{journal}{\aap} \bibinfo{volume}{666}, \bibinfo{pages}{A124}.
\newblock \DOIprefix\doi{10.1051/0004-6361/202244323}, \href{http://arxiv.org/abs/2207.10921}{{\tt arXiv:2207.10921}}.
\bibitem[{{Aharonian} et~al.(2019){Aharonian}, {Yang} and {de O{\~n}a Wilhelmi}}]{Aharonian1}
\bibinfo{author}{{Aharonian}, F.}, \bibinfo{author}{{Yang}, R.}, \bibinfo{author}{{de O{\~n}a Wilhelmi}, E.}, \bibinfo{year}{2019}.
\newblock \bibinfo{title}{{Massive stars as major factories of Galactic cosmic rays}}.
\newblock \bibinfo{journal}{Nature Astronomy} \bibinfo{volume}{3}, \bibinfo{pages}{561--567}.
\newblock \DOIprefix\doi{10.1038/s41550-019-0724-0}, \href{http://arxiv.org/abs/1804.02331}{{\tt arXiv:1804.02331}}.
\bibitem[{{Baume} et~al.(2004){Baume}, {V{\'a}zquez} and {Carraro}}]{markarian}
\bibinfo{author}{{Baume}, G.}, \bibinfo{author}{{V{\'a}zquez}, R.A.}, \bibinfo{author}{{Carraro}, G.}, \bibinfo{year}{2004}.
\newblock \bibinfo{title}{{The young open cluster Markarian 50}}.
\newblock \bibinfo{journal}{\mnras} \bibinfo{volume}{355}, \bibinfo{pages}{475--484}.
\newblock \DOIprefix\doi{10.1111/j.1365-2966.2004.08337.x}.
\bibitem[{{Beasor} et~al.(2021){Beasor}, {Davies} and {Smith et al.}}]{wd1age}
\bibinfo{author}{{Beasor}, E.R.}, \bibinfo{author}{{Davies}, B.}, \bibinfo{author}{{Smith et al.}, N.}, \bibinfo{year}{2021}.
\newblock \bibinfo{title}{{The Age of Westerlund 1 Revisited}}.
\newblock \bibinfo{journal}{\apj} \bibinfo{volume}{912}, \bibinfo{pages}{16}.
\newblock \DOIprefix\doi{10.3847/1538-4357/abec44}, \href{http://arxiv.org/abs/2103.02609}{{\tt arXiv:2103.02609}}.
\bibitem[{{Berlanas} et~al.(2018){Berlanas}, {Herrero} and {Comer{\'o}n et al.}}]{Cygrot}
\bibinfo{author}{{Berlanas}, S.R.}, \bibinfo{author}{{Herrero}, A.}, \bibinfo{author}{{Comer{\'o}n et al.}, F.}, \bibinfo{year}{2018}.
\newblock \bibinfo{title}{{New massive members of Cygnus OB2}}.
\newblock \bibinfo{journal}{\aap} \bibinfo{volume}{612}, \bibinfo{pages}{A50}.
\newblock \DOIprefix\doi{10.1051/0004-6361/201731856}, \href{http://arxiv.org/abs/1711.06945}{{\tt arXiv:1711.06945}}.
\bibitem[{{Berlanas} et~al.(2019){Berlanas}, {Wright} and {Herrero et al.}}]{2019MNRAS.484.1838B}
\bibinfo{author}{{Berlanas}, S.R.}, \bibinfo{author}{{Wright}, N.J.}, \bibinfo{author}{{Herrero et al.}, A.}, \bibinfo{year}{2019}.
\newblock \bibinfo{title}{{Disentangling the spatial substructure of Cygnus OB2 from Gaia DR2}}.
\newblock \bibinfo{journal}{\mnras} \bibinfo{volume}{484}, \bibinfo{pages}{1838--1842}.
\newblock \DOIprefix\doi{10.1093/mnras/stz117}, \href{http://arxiv.org/abs/1901.02959}{{\tt arXiv:1901.02959}}.
\bibitem[{{Cant{\'o}} et~al.(2000){Cant{\'o}}, {Raga} and {Rodr{\'\i}guez}}]{canto_morlino}
\bibinfo{author}{{Cant{\'o}}, J.}, \bibinfo{author}{{Raga}, A.C.}, \bibinfo{author}{{Rodr{\'\i}guez}, L.F.}, \bibinfo{year}{2000}.
\newblock \bibinfo{title}{{The Hot, Diffuse Gas in a Dense Cluster of Massive Stars}}.
\newblock \bibinfo{journal}{\apj} \bibinfo{volume}{536}, \bibinfo{pages}{896--901}.
\newblock \DOIprefix\doi{10.1086/308983}.
\bibitem[{{Cao} et~al.(2024){Cao}, {Aharonian} and {An et al.}}]{lhaaso_cat}
\bibinfo{author}{{Cao}, Z.}, \bibinfo{author}{{Aharonian}, F.}, \bibinfo{author}{{An et al.}, Q.}, \bibinfo{year}{2024}.
\newblock \bibinfo{title}{{The First LHAASO Catalog of Gamma-Ray Sources}}.
\newblock \bibinfo{journal}{\apjs} \bibinfo{volume}{271}, \bibinfo{pages}{25}.
\newblock \DOIprefix\doi{10.3847/1538-4365/acfd29}, \href{http://arxiv.org/abs/2305.17030}{{\tt arXiv:2305.17030}}.
\bibitem[{{Celli} et~al.(2024){Celli}, {Specovius} and {Menchiari et al.}}]{celli}
\bibinfo{author}{{Celli}, S.}, \bibinfo{author}{{Specovius}, A.}, \bibinfo{author}{{Menchiari et al.}, S.}, \bibinfo{year}{2024}.
\newblock \bibinfo{title}{{Mass and wind luminosity of young Galactic open clusters in Gaia DR2}}.
\newblock \bibinfo{journal}{\aap} \bibinfo{volume}{686}, \bibinfo{pages}{A118}.
\newblock \DOIprefix\doi{10.1051/0004-6361/202348541}, \href{http://arxiv.org/abs/2311.09089}{{\tt arXiv:2311.09089}}.
\bibitem[{{Cherenkov Telescope Array Consortium} et~al.(2019){Cherenkov Telescope Array Consortium}, {Acharya}, {Agudo} and {Al Samarai et al.}}]{ctao}
\bibinfo{author}{{Cherenkov Telescope Array Consortium}}, \bibinfo{author}{{Acharya}, B.S.}, \bibinfo{author}{{Agudo}, I.}, \bibinfo{author}{{Al Samarai et al.}, I.}, \bibinfo{year}{2019}.
\newblock \bibinfo{title}{{Science with the Cherenkov Telescope Array}}.
\newblock \DOIprefix\doi{10.1142/10986}.
\bibitem[{{Comer{\'o}n} and {Pasquali}(2007)}]{cyg16}
\bibinfo{author}{{Comer{\'o}n}, F.}, \bibinfo{author}{{Pasquali}, A.}, \bibinfo{year}{2007}.
\newblock \bibinfo{title}{{A very massive runaway star from Cygnus OB2}}.
\newblock \bibinfo{journal}{\aap} \bibinfo{volume}{467}, \bibinfo{pages}{L23--L27}.
\newblock \DOIprefix\doi{10.1051/0004-6361:20077304}, \href{http://arxiv.org/abs/0704.0676}{{\tt arXiv:0704.0676}}.
\bibitem[{{Comer{\'o}n} and {Pasquali}(2012)}]{Cyg10myr}
\bibinfo{author}{{Comer{\'o}n}, F.}, \bibinfo{author}{{Pasquali}, A.}, \bibinfo{year}{2012}.
\newblock \bibinfo{title}{{New members of the massive stellar population in Cygnus}}.
\newblock \bibinfo{journal}{\aap} \bibinfo{volume}{543}, \bibinfo{pages}{A101}.
\newblock \DOIprefix\doi{10.1051/0004-6361/201219022}.
\bibitem[{{Davies} et~al.(2012){Davies}, {Clark} and {Trombley et al.}}]{danks}
\bibinfo{author}{{Davies}, B.}, \bibinfo{author}{{Clark}, J.S.}, \bibinfo{author}{{Trombley et al.}, C.}, \bibinfo{year}{2012}.
\newblock \bibinfo{title}{{The G305 star-forming complex: the central star clusters Danks 1 and Danks 2}}.
\newblock \bibinfo{journal}{\mnras} \bibinfo{volume}{419}, \bibinfo{pages}{1871--1886}.
\newblock \DOIprefix\doi{10.1111/j.1365-2966.2011.19736.x}, \href{http://arxiv.org/abs/1109.0314}{{\tt arXiv:1109.0314}}.
\bibitem[{{Donath} et~al.(2023){Donath}, {Terrier} and {Remy et al.}}]{gammapy:2023}
\bibinfo{author}{{Donath}, A.}, \bibinfo{author}{{Terrier}, R.}, \bibinfo{author}{{Remy et al.}, Q.}, \bibinfo{year}{2023}.
\newblock \bibinfo{title}{{Gammapy: A Python package for gamma-ray astronomy}}.
\newblock \bibinfo{journal}{\aap} \bibinfo{volume}{678}, \bibinfo{pages}{A157}.
\newblock \DOIprefix\doi{10.1051/0004-6361/202346488}, \href{http://arxiv.org/abs/2308.13584}{{\tt arXiv:2308.13584}}.
\bibitem[{{Drew} et~al.(2008){Drew}, {Greimel} and {Irwin et al.}}]{cygage}
\bibinfo{author}{{Drew}, J.E.}, \bibinfo{author}{{Greimel}, R.}, \bibinfo{author}{{Irwin et al.}, M.J.}, \bibinfo{year}{2008}.
\newblock \bibinfo{title}{{Early-A stars from IPHAS, and their distribution in and around the Cyg OB2 association}}.
\newblock \bibinfo{journal}{\mnras} \bibinfo{volume}{386}, \bibinfo{pages}{1761--1773}.
\newblock \DOIprefix\doi{10.1111/j.1365-2966.2008.13147.x}, \href{http://arxiv.org/abs/0802.3868}{{\tt arXiv:0802.3868}}.
\bibitem[{{Gabici}(2023)}]{gabici}
\bibinfo{author}{{Gabici}, S.}, \bibinfo{year}{2023}.
\newblock \bibinfo{title}{{Cosmic rays from star clusters}}.
\newblock \bibinfo{journal}{arXiv e-prints} , \bibinfo{pages}{arXiv:2307.01596}\DOIprefix\doi{10.48550/arXiv.2307.01596}, \href{http://arxiv.org/abs/2307.01596}{{\tt arXiv:2307.01596}}.
\bibitem[{Gabici et~al.(2019)Gabici, Evoli, Gaggero, Lipari, Mertsch, Orlando, Strong and Vittino}]{CR_review}
\bibinfo{author}{Gabici, S.}, \bibinfo{author}{Evoli, C.}, \bibinfo{author}{Gaggero, D.}, \bibinfo{author}{Lipari, P.}, \bibinfo{author}{Mertsch, P.}, \bibinfo{author}{Orlando, E.}, \bibinfo{author}{Strong, A.}, \bibinfo{author}{Vittino, A.}, \bibinfo{year}{2019}.
\newblock \bibinfo{title}{The origin of galactic cosmic rays: Challenges to the standard paradigm}.
\newblock \bibinfo{journal}{International Journal of Modern Physics D} \bibinfo{volume}{28}, \bibinfo{pages}{1930022}.
\newblock \URLprefix \url{https://doi.org/10.1142/S0218271819300222}, \DOIprefix\doi{10.1142/S0218271819300222}, \href{http://arxiv.org/abs/https://doi.org/10.1142/S0218271819300222}{{\tt arXiv:https://doi.org/10.1142/S0218271819300222}}.
\bibitem[{{Gelmini}(2009)}]{Gab}
\bibinfo{author}{{Gelmini}, G.B.}, \bibinfo{year}{2009}.
\newblock \bibinfo{title}{{High energy cosmic rays}}, in: \bibinfo{booktitle}{Journal of Physics Conference Series}, \bibinfo{publisher}{IOP}. p. \bibinfo{pages}{012012}.
\newblock \DOIprefix\doi{10.1088/1742-6596/171/1/012012}, \href{http://arxiv.org/abs/0903.4716}{{\tt arXiv:0903.4716}}.
\bibitem[{{Gennaro} et~al.(2011){Gennaro}, {Brandner} and {Stolte et al.}}]{wd1agelow}
\bibinfo{author}{{Gennaro}, M.}, \bibinfo{author}{{Brandner}, W.}, \bibinfo{author}{{Stolte et al.}, A.}, \bibinfo{year}{2011}.
\newblock \bibinfo{title}{{Mass segregation and elongation of the starburst cluster Westerlund 1}}.
\newblock \bibinfo{journal}{\mnras} \bibinfo{volume}{412}, \bibinfo{pages}{2469--2488}.
\newblock \DOIprefix\doi{10.1111/j.1365-2966.2010.18068.x}, \href{http://arxiv.org/abs/1011.5223}{{\tt arXiv:1011.5223}}.
\bibitem[{{Guarcello} et~al.(2024){Guarcello}, {Almendros-Abad} and {Lovell et al.}}]{Wd1_2}
\bibinfo{author}{{Guarcello}, M.G.}, \bibinfo{author}{{Almendros-Abad}, V.}, \bibinfo{author}{{Lovell et al.}, J.B.}, \bibinfo{year}{2024}.
\newblock \bibinfo{title}{{EWOCS-III: JWST observations of the supermassive star cluster Westerlund 1}}.
\newblock \bibinfo{journal}{arXiv e-prints} , \bibinfo{pages}{arXiv:2411.13051}\DOIprefix\doi{10.48550/arXiv.2411.13051}, \href{http://arxiv.org/abs/2411.13051}{{\tt arXiv:2411.13051}}.
\bibitem[{{Gupta} et~al.(2021){Gupta}, {Jose} and {More et al.}}]{Cyg5myr}
\bibinfo{author}{{Gupta}, S.}, \bibinfo{author}{{Jose}, J.}, \bibinfo{author}{{More et al.}, S.}, \bibinfo{year}{2021}.
\newblock \bibinfo{title}{{Subaru Hyper Suprime-Cam Survey of Cygnus OB2 Complex - I. Introduction, photometry, and source catalogue}}.
\newblock \bibinfo{journal}{\mnras} \bibinfo{volume}{508}, \bibinfo{pages}{3388--3407}.
\newblock \DOIprefix\doi{10.1093/mnras/stab2751}, \href{http://arxiv.org/abs/2109.11009}{{\tt arXiv:2109.11009}}.
\bibitem[{{H.~E.~S.~S. Collaboration} et~al.(2018){H.~E.~S.~S. Collaboration}, {Abdalla}, {Abramowski} and {Aharonian et al.}}]{hess_morlino}
\bibinfo{author}{{H.~E.~S.~S. Collaboration}}, \bibinfo{author}{{Abdalla}, H.}, \bibinfo{author}{{Abramowski}, A.}, \bibinfo{author}{{Aharonian et al.}, F.}, \bibinfo{year}{2018}.
\newblock \bibinfo{title}{{The H.E.S.S. Galactic plane survey}}.
\newblock \bibinfo{journal}{\aap} \bibinfo{volume}{612}, \bibinfo{pages}{A1}.
\newblock \DOIprefix\doi{10.1051/0004-6361/201732098}, \href{http://arxiv.org/abs/1804.02432}{{\tt arXiv:1804.02432}}.
\bibitem[{{Hanson}(2003)}]{cygagelow}
\bibinfo{author}{{Hanson}, M.M.}, \bibinfo{year}{2003}.
\newblock \bibinfo{title}{{A Study of Cygnus OB2: Pointing the Way toward Finding Our Galaxy's Super-Star Clusters}}.
\newblock \bibinfo{journal}{\apj} \bibinfo{volume}{597}, \bibinfo{pages}{957--969}.
\newblock \DOIprefix\doi{10.1086/378508}, \href{http://arxiv.org/abs/astro-ph/0307540}{{\tt arXiv:astro-ph/0307540}}.
\bibitem[{{H{\"a}rer} et~al.(2023){H{\"a}rer}, {Reville} and {Hinton et al.}}]{Wd1}
\bibinfo{author}{{H{\"a}rer}, L.K.}, \bibinfo{author}{{Reville}, B.}, \bibinfo{author}{{Hinton et al.}, J.}, \bibinfo{year}{2023}.
\newblock \bibinfo{title}{{Understanding the TeV {\ensuremath{\gamma}}-ray emission surrounding the young massive star cluster Westerlund 1}}.
\newblock \bibinfo{journal}{\aap} \bibinfo{volume}{671}, \bibinfo{pages}{A4}.
\newblock \DOIprefix\doi{10.1051/0004-6361/202245444}, \href{http://arxiv.org/abs/2301.10496}{{\tt arXiv:2301.10496}}.
\bibitem[{{Hillas}(1984)}]{Hillas}
\bibinfo{author}{{Hillas}, A.M.}, \bibinfo{year}{1984}.
\newblock \bibinfo{title}{{The Origin of Ultra-High-Energy Cosmic Rays}}.
\newblock \bibinfo{journal}{\araa} \bibinfo{volume}{22}, \bibinfo{pages}{425--444}.
\newblock \DOIprefix\doi{10.1146/annurev.aa.22.090184.002233}.
\bibitem[{{Koo} and {McKee}(1992)}]{koo}
\bibinfo{author}{{Koo}, B.C.}, \bibinfo{author}{{McKee}, C.F.}, \bibinfo{year}{1992}.
\newblock \bibinfo{title}{{Dynamics of Wind Bubbles and Superbubbles. I. Slow Winds and Fast Winds}}.
\newblock \bibinfo{journal}{\apj} \bibinfo{volume}{388}, \bibinfo{pages}{93}.
\newblock \DOIprefix\doi{10.1086/171132}.
\bibitem[{{Lhaaso Collaboration}(2024)}]{CygLHAASO}
\bibinfo{author}{{Lhaaso Collaboration}}, \bibinfo{year}{2024}.
\newblock \bibinfo{title}{{An ultrahigh-energy {\ensuremath{\gamma}} -ray bubble powered by a super PeVatron}}.
\newblock \bibinfo{journal}{Science Bulletin} \bibinfo{volume}{69}, \bibinfo{pages}{449--457}.
\newblock \DOIprefix\doi{10.1016/j.scib.2023.12.040}, \href{http://arxiv.org/abs/2310.10100}{{\tt arXiv:2310.10100}}.
\bibitem[{{Liu} et~al.(2024){Liu}, {Liu} and {Yang}}]{Danks_fermi}
\bibinfo{author}{{Liu}, J.h.}, \bibinfo{author}{{Liu}, B.}, \bibinfo{author}{{Yang}, R.z.}, \bibinfo{year}{2024}.
\newblock \bibinfo{title}{{Detection of extended gamma-ray emission in the vicinity of Cl Danks 1 and 2}}.
\newblock \bibinfo{journal}{\mnras} \bibinfo{volume}{535}, \bibinfo{pages}{1526--1532}.
\newblock \DOIprefix\doi{10.1093/mnras/stae2404}, \href{http://arxiv.org/abs/2406.03320}{{\tt arXiv:2406.03320}}.
\bibitem[{{Mitchell} et~al.(2024){Mitchell}, {Morlino} and {Celli et al.}}]{mitchell}
\bibinfo{author}{{Mitchell}, A.M.W.}, \bibinfo{author}{{Morlino}, G.}, \bibinfo{author}{{Celli et al.}, S.}, \bibinfo{year}{2024}.
\newblock \bibinfo{title}{{Probing Stellar Clusters from Gaia DR2 as Galactic PeVatrons: I -- Expected Gamma-ray and Neutrino Emission}}.
\newblock \bibinfo{journal}{arXiv e-prints} , \bibinfo{pages}{arXiv:2403.16650}\DOIprefix\doi{10.48550/arXiv.2403.16650}, \href{http://arxiv.org/abs/2403.16650}{{\tt arXiv:2403.16650}}.
\bibitem[{{Morlino} et~al.(2021){Morlino}, {Blasi} and {Peretti et al.}}]{morlino}
\bibinfo{author}{{Morlino}, G.}, \bibinfo{author}{{Blasi}, P.}, \bibinfo{author}{{Peretti et al.}, E.}, \bibinfo{year}{2021}.
\newblock \bibinfo{title}{{Particle acceleration in winds of star clusters}}.
\newblock \bibinfo{journal}{\mnras} \bibinfo{volume}{504}, \bibinfo{pages}{6096--6105}.
\newblock \DOIprefix\doi{10.1093/mnras/stab690}, \href{http://arxiv.org/abs/2102.09217}{{\tt arXiv:2102.09217}}.
\bibitem[{Observatory and Consortium(2021)}]{ctairf}
\bibinfo{author}{Observatory, C.T.A.}, \bibinfo{author}{Consortium, C.T.A.}, \bibinfo{year}{2021}.
\newblock \bibinfo{title}{Ctao instrument response functions - prod5 version v0.1}.
\newblock \URLprefix \url{https://doi.org/10.5281/zenodo.5499840}, \DOIprefix\doi{10.5281/zenodo.5499840}.
\bibitem[{{Orellana} et~al.(2021){Orellana}, {De Biasi} and {Pa{\'\i}z}}]{2021MNRAS.502.6080O}
\bibinfo{author}{{Orellana}, R.B.}, \bibinfo{author}{{De Biasi}, M.S.}, \bibinfo{author}{{Pa{\'\i}z}, L.G.}, \bibinfo{year}{2021}.
\newblock \bibinfo{title}{{New members of Cygnus OB2 from Gaia DR2}}.
\newblock \bibinfo{journal}{\mnras} \bibinfo{volume}{502}, \bibinfo{pages}{6080--6093}.
\newblock \DOIprefix\doi{10.1093/mnras/stab457}.
\bibitem[{Project(2022)}]{astriirf}
\bibinfo{author}{Project, A.}, \bibinfo{year}{2022}.
\newblock \bibinfo{title}{Astri mini-array instrument response functions (prod2, v1.0)}.
\newblock \URLprefix \url{https://doi.org/10.5281/zenodo.6827882}, \DOIprefix\doi{10.5281/zenodo.6827882}.
\bibitem[{{Scuderi} et~al.(2022){Scuderi}, {Giuliani} and {Pareschi et al.}}]{astri}
\bibinfo{author}{{Scuderi}, S.}, \bibinfo{author}{{Giuliani}, A.}, \bibinfo{author}{{Pareschi et al.}, G.}, \bibinfo{year}{2022}.
\newblock \bibinfo{title}{{The ASTRI Mini-Array of Cherenkov telescopes at the Observatorio del Teide}}.
\newblock \bibinfo{journal}{Journal of High Energy Astrophysics} \bibinfo{volume}{35}, \bibinfo{pages}{52--68}.
\newblock \DOIprefix\doi{10.1016/j.jheap.2022.05.001}, \href{http://arxiv.org/abs/2208.04571}{{\tt arXiv:2208.04571}}.
\bibitem[{{Vieu} et~al.(2024){Vieu}, {Larkin} and {H{\"a}rer et al.}}]{vieu_morlino}
\bibinfo{author}{{Vieu}, T.}, \bibinfo{author}{{Larkin}, C.J.K.}, \bibinfo{author}{{H{\"a}rer et al.}, L.}, \bibinfo{year}{2024}.
\newblock \bibinfo{title}{{Hydrodynamic simulation of Cygnus OB2: the absence of a cluster wind termination shock}}.
\newblock \bibinfo{journal}{\mnras} \bibinfo{volume}{532}, \bibinfo{pages}{2174--2188}.
\newblock \DOIprefix\doi{10.1093/mnras/stae1627}, \href{http://arxiv.org/abs/2406.13589}{{\tt arXiv:2406.13589}}.
\bibitem[{{Yadav} et~al.(2017){Yadav}, {Mukherjee} and {Sharma et al.}}]{eta_morlino}
\bibinfo{author}{{Yadav}, N.}, \bibinfo{author}{{Mukherjee}, D.}, \bibinfo{author}{{Sharma et al.}, P.}, \bibinfo{year}{2017}.
\newblock \bibinfo{title}{{How multiple supernovae overlap to form superbubbles}}.
\newblock \bibinfo{journal}{\mnras} \bibinfo{volume}{465}, \bibinfo{pages}{1720--1740}.
\newblock \DOIprefix\doi{10.1093/mnras/stw2522}, \href{http://arxiv.org/abs/1603.00815}{{\tt arXiv:1603.00815}}.
\bibitem[{{Zabalza}(2015)}]{naima}
\bibinfo{author}{{Zabalza}, V.}, \bibinfo{year}{2015}.
\newblock \bibinfo{title}{{Naima: a Python package for inference of particle distribution properties from nonthermal spectra}}, in: \bibinfo{booktitle}{34th International Cosmic Ray Conference (ICRC2015)}, p. \bibinfo{pages}{922}.
\newblock \DOIprefix\doi{10.22323/1.236.0922}, \href{http://arxiv.org/abs/1509.03319}{{\tt arXiv:1509.03319}}.

\end{thebibliography}

\end{document}